\documentclass{article}
\usepackage{coopshift_preprint,times}
\usepackage[T1]{fontenc}
\usepackage[utf8]{inputenc}
\usepackage{microtype,amsmath,amssymb,booktabs,longtable,tabularx,array,graphicx,xcolor}
\usepackage{hyperref,url,xurl,seqsplit,placeins}
\hypersetup{hidelinks,pdftitle={CoopShift: Post-Training Amplifies the Memory Curse in LLM Cooperation},pdfsubject={Cooperation across released training stages},pdfkeywords={large language models; post-training; multi-agent cooperation; repeated games; interaction history; behavioral evaluation}}

\expandafter\newcommand\csname csmodeliconai2\endcsname{\includegraphics[width=1.05em,height=1.05em,keepaspectratio]{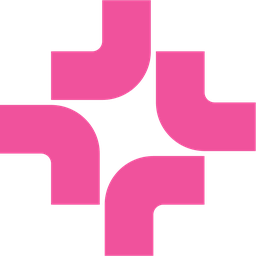}}
\expandafter\newcommand\csname csmodelicondeepseek\endcsname{\includegraphics[width=1.05em,height=1.05em,keepaspectratio]{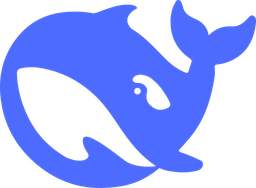}}
\expandafter\newcommand\csname csmodeliconhuggingface\endcsname{\includegraphics[width=1.05em,height=1.05em,keepaspectratio]{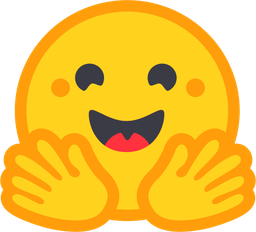}}
\expandafter\newcommand\csname csmodeliconinternlm\endcsname{\includegraphics[width=1.05em,height=1.05em,keepaspectratio]{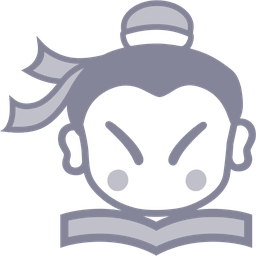}}
\expandafter\newcommand\csname csmodeliconnvidia\endcsname{\includegraphics[width=1.05em,height=1.05em,keepaspectratio]{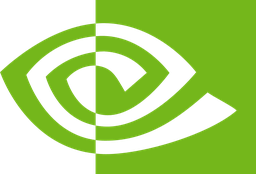}}
\expandafter\newcommand\csname csmodelicontulu\endcsname{\includegraphics[width=1.05em,height=1.05em,keepaspectratio]{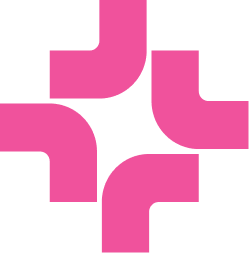}}
\expandafter\newcommand\csname csmodeliconzephyr\endcsname{\includegraphics[width=1.05em,height=1.05em,keepaspectratio]{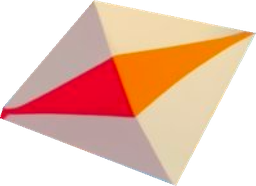}}
\newcommand{\csmodel}[2]{%
  \makebox[1.15em][c]{\raisebox{-0.17em}{\csname csmodelicon#1\endcsname}}%
  \hspace{0.3em}#2}

\title{CoopShift: Post-Training Amplifies the\\Memory Curse in LLM Cooperation}

\author{%
\begin{tabular}[t]{@{}l@{}}
\rule{0pt}{24pt}\textbf{Jiayuan Liu}$^{1,2}$ \quad \textbf{Shiyi Du}$^{1}$ \quad \textbf{Tianqin Li}$^{1}$ \\
\textbf{Emanuel Tewolde}$^{1,2}$ \quad \textbf{Keyang Xuan}$^{3}$ \quad \textbf{Vincent Conitzer}$^{1,2}$\thanks{Corresponding author.} \\[6pt]
\small $^{1}$Carnegie Mellon University \quad $^{2}$Foundations of Cooperative AI Lab (FOCAL) \\
\small $^{3}$The University of Texas at Austin \\[4pt]
\small\texttt{\{jiayuan4,shiyid,tianqinl,etewolde\}@cs.cmu.edu}\quad\texttt{kx932@eid.utexas.edu} \\
\small\href{mailto:conitzer@cs.cmu.edu}{\texttt{conitzer@cs.cmu.edu}}
\end{tabular}%
}
\date{}
\hypersetup{pdfauthor={Jiayuan Liu, Shiyi Du, Tianqin Li, Emanuel Tewolde, Keyang Xuan, Vincent Conitzer}}

\begin{document}
\maketitle
\begin{abstract}
How does post-training change cooperation during repeated interaction? Within each of seven released model lineages, we compare pretrained (Base), supervised fine-tuned (SFT), and subsequent preference- or reward-optimized (Aligned) checkpoints across seven social dilemmas. Copies of each checkpoint play one another for 200 rounds, with visible history limited to the last two or forty rounds. The memory curse refers to lower cooperation when agents can see more past rounds. Our central finding is that post-training amplifies this decline. Averaged across the six binary-action games, expanding visible history reduces cooperation at every SFT and Aligned checkpoint. At each post-trained stage, these declines exceed Base's in six of seven lineages. Most of this average amplification is already present at SFT. With long history, both post-trained stages also cooperate less than Base on average in each binary-action game. In the symmetric binary games, their fitted cooperation rates fall more steeply on average than Base's as per-round cooperation costs rise, benefits fall, or noncooperation becomes more rewarding. The difference in cooperation rates between Aligned and SFT also changes from one-shot decisions to repeated play. When given the same past records, their choices depend on whether the instructions describe a final round or possible future interaction. Post-training evaluation should therefore go beyond isolated choices to assess whether cooperation is sustained in repeated interaction and how it depends on payoffs and visible history.
\end{abstract}

\section{Introduction}
\label{sec:introduction}
Post-training adapts models to follow instructions and satisfy evaluated preferences \citep{ouyang2022training,rafailov2023dpo}. Repeated interaction adds a challenge: sustaining cooperation despite incentives to deviate \citep{piatti2024cooperate}. In a public-goods game (PGG), for example, players keep their endowments or contribute them to a pool. The pool is multiplied by a factor greater than one and shared equally among all players, including noncontributors. In our PGG, mutual contribution maximizes group payoff and gives everyone more than mutual retention. Yet withholding one's contribution while others contribute earns that player even more this round, so mutual contribution is not a one-round Nash equilibrium. Repetition changes the trade-off: the immediate gain from defection can be outweighed by losing future mutual cooperation \citep{fudenberg1986folk}. We ask: \emph{how does cooperation differ across training stages, and how do these differences change when agents can see more of their interaction history?}
\begin{figure}[!t]
\centering
\includegraphics[width=\linewidth]{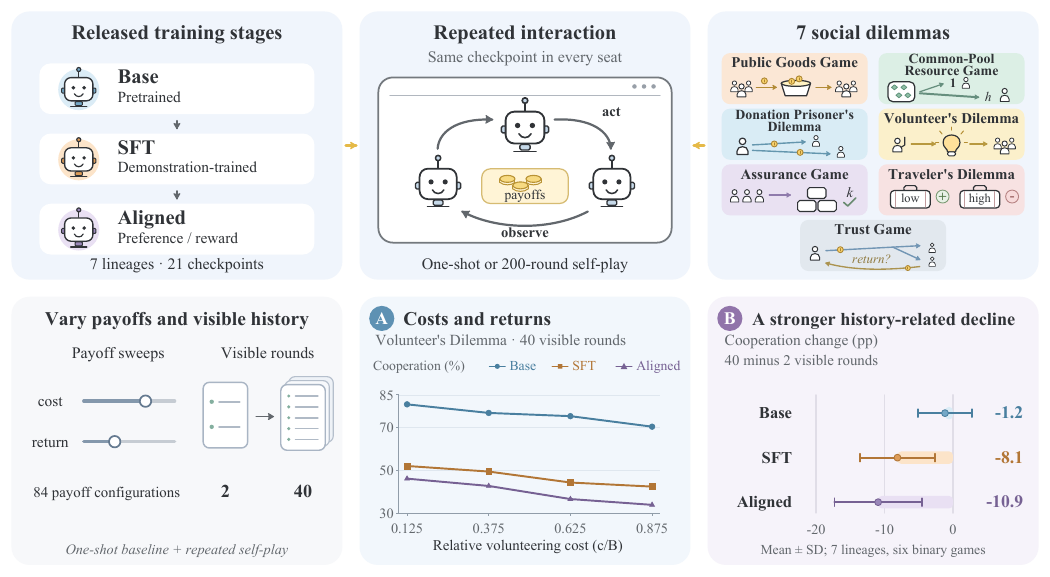}
\caption{\textbf{CoopShift: cooperation across released post-training stages.} Seven lineages (21 checkpoints) play seven social dilemmas across 84 payoff configurations. One-shot decisions complement 200-round self-play. \textbf{A:} Volunteer's Dilemma with a forty-round history limit (HL40): mean cooperation across seven lineages versus relative volunteering cost $c/B$, where $c$ is cost and $B$ each player's benefit. The y axis spans 30--85\%. \textbf{B:} Forty-minus-two-round history responses, in percentage points (pp): mean $\pm$ standard deviation (SD) across seven lineage summaries, each averaging the six binary games (Table~\ref{tab:history-headline}).}
\label{fig:teaser}
\end{figure}

Past interaction matters because agents can use it to decide whether to keep cooperating, withdraw, or try to restore cooperation. A longer record gives them more evidence about their partners; its effect on cooperation depends on how that evidence changes their choices. Human experiments separate private contribution costs from public returns \citep{goeree2002private}, reveal payoff-dependent conditional cooperation \citep{baader2026conditional}, and link declining contributions to imperfect conditional cooperation \citep{fischbacher2010social}. In human experiments, intermediate visible histories can also support more cooperation than longer records \citep{ma2021limited}.

Language-model cooperation depends on objectives, partners, and repeated interaction \citep{phelps2023machine,akata2024repeated}. \citet{liu2026memorycurse} document a \emph{memory curse}: agents can cooperate less when their prompts contain more past rounds. Here, \emph{memory} means the interaction record visible at each decision. Within each model family, or \emph{lineage}, we compare pretrained Base, demonstration-trained SFT, and a subsequent preference- or reward-optimized Aligned checkpoint. We ask whether expanding visible history produces a larger cooperation decline at the post-trained checkpoints than at Base. This larger short-to-long-history decline is what we call \emph{amplification} of the memory curse.

We evaluate 21 checkpoints from seven lineages in seven social dilemmas: PGG, a common-pool resource game (CPR), Donation Prisoner's Dilemma (Donation-PD), the Volunteer's Dilemma (Volunteer), Assurance, Traveler's Dilemma (TD), and the Trust game (Table~\ref{tab:game-economic-guide}). One-shot prompts present a single decision without past interaction or future-round statements. In repeated play, copies of one checkpoint interact for 200 recorded rounds, seeing up to the two or forty most recent rounds. Separate tests give different checkpoints the same past-interaction records, either constructed by the researchers or saved from earlier games, and compare their current choices.

For each binary game, the cooperation rate records how often its designated cooperative action is chosen. At a fixed PGG or Donation-PD payoff setting, this rate also determines the group's average earned points. We use \emph{collective welfare} to mean the sum of all players' earned points.

Our central finding is that \textbf{post-training amplifies the memory curse (Section~\ref{sec:f2}).} After averaging the six binary games equally within each lineage, increasing visible history from two to forty rounds lowers cooperation at every SFT and Aligned checkpoint, by about 8 and 11 percentage points (pp) on average, respectively. Base's mean decline is about 1 pp, and its cooperation increases in five of seven lineages. At each post-trained stage, the decline exceeds Base's in six of seven lineages, and most of the average amplification is already present at SFT. We extend prior memory evidence by showing where the response to longer histories becomes stronger along released training sequences.

With forty visible rounds, SFT and Aligned cooperate less than Base on average in each binary game. \textbf{Their cooperation rates also fall more than Base's, on average, as per-round cooperation costs rise, benefits fall, or noncooperation becomes more rewarding (Section~\ref{sec:f1}).} This comparison uses fitted trends across each game's tested payoff values and holds in the symmetric binary games. TD has a different outcome: both stages choose higher mean claims than Base. Lower contributions translate into lower mean PGG and Donation-PD payoffs. Differences in cooperation between stages therefore depend on both payoffs and visible history. Section~\ref{sec:context} compares one-shot with repeated play, then gives checkpoints the same past records but different future instructions.

\section{Experimental Design}
\label{sec:design}

We compare 21 checkpoints from seven open-weight lineages. Aligned objectives correspond to direct preference optimization (DPO), reinforcement learning (RL) from human feedback (RLHF), or group relative policy optimization (GRPO). In Base / SFT / Aligned order, the display names are: \mbox{\csmodel{ai2}{OLMo-3-7B}} (Base / Instruct-SFT / Instruct-DPO); \mbox{\csmodel{huggingface}{SmolLM2-1.7B}} (Base / SFT-only / Instruct); \mbox{\csmodel{zephyr}{Mistral/Zephyr-7B}} (Mistral-v0.1 / Mistral-SFT-Beta / Zephyr-Beta); \mbox{\csmodel{tulu}{Llama/T\"ulu-3-8B}} (Llama-3.1 / T\"ulu-3-SFT / T\"ulu-3-DPO); \mbox{\csmodel{internlm}{InternLM2-7B}} (Base / Chat-SFT / Chat); \mbox{\csmodel{deepseek}{DeepSeekMath-7B}} (Base / Instruct / RL); and \mbox{\csmodel{nvidia}{Nemotron-Cascade-8B}} (Qwen3-Base / Cascade-SFT / Cascade-RLHF). We abbreviate stages as B, S, A. Appendix~\ref{app:models} identifies the exact released checkpoints.

A \emph{payoff configuration} fixes a game's player count and all its payoff parameters. In the main one-shot and repeated-game evaluation, every checkpoint uses the same 84 configurations: PGG (8), CPR (16), Donation-PD (12), Volunteer (12), Assurance (8), TD (16), and Trust (12). Each is used in one-shot, HL2, and HL40 evaluations. In the first five games, players have the same two action options and payoff rule; TD offers multiple claims; Trust has an investor and two trustees. Table~\ref{tab:game-economic-guide} explains the games and parameters; Appendix~\ref{app:games} gives their complete payoff rules.
\begin{table}[!htbp]
\centering\small
\caption{What the games measure. The main parameter changes a cost, return, or coordination reward, depending on the game.}
\label{tab:game-economic-guide}
\fontsize{8.5}{9.8}\selectfont
\begin{tabularx}{\linewidth}{@{}p{0.13\linewidth}p{0.28\linewidth}X@{}}
\toprule
Game & Decision and tension & Meaning of the raw main parameter\\
\midrule
PGG & Contribute a unit to a shared pool or retain it.
& The marginal per-capita return (MPCR), $\alpha=r/n$, is each player's return per contributed unit, where $r$ is the pool multiplier and $n$ the group size. Higher $\alpha$ lowers private cost $1-\alpha$.\\
CPR & Restrain extraction or extract more; each unit earns less as total extraction rises.
& Low extraction is 1 and high extraction is $h>1$. Higher $\theta$ increases high extraction's private advantage at fixed $h$; payoffs depend on current-round extraction.\\
Donation-PD & Pay $c$ to give $b$ to each other player, or give nothing.
& $c$ is the donor's cost and $b$ the benefit to each of two others. Higher $R=2b/c$, at fixed $c$, raises external returns; private cost stays $c$.\\
Volunteer & Supply a shared benefit or rely on another volunteer; one suffices.
& Each player receives $B$ if anyone volunteers; every volunteer pays $c$, so extra volunteers duplicate costs. Higher $\rho=c/B$, at fixed $B$, increases volunteering cost.\\
Assurance & Join a threshold project or take a sure outside payoff.
& A successful contributor earns a fraction $\lambda$ more than the sure outside payoff. Success requires at least $k$ contributors, including itself; unsuccessful contributors receive zero. Higher $\lambda$ raises the reward for successful coordination.\\
TD & Claim an amount; the lower claimant gets a bonus, the higher a penalty.
& $K$ is the number of legal claim levels. Higher $\tau$, at fixed $K$, raises bonus/penalty and one-step undercutting gain. Outcome: claim normalized to $[0,1]$.\\
Trust & Choose whether to invest; two trustees state simultaneous, contingent return intentions.
& Investment gives each trustee $5m$, where $m$ is the multiplier. A returning trustee sends back $5(1+s)$; higher $s$ raises the investor's return and the trustee's cost.\\
\bottomrule
\end{tabularx}
\end{table}

\textbf{Whose cooperation is measured?} Within each run, we average players and then rounds. We then average runs and payoff configurations. Thus Trust counts each of its three players equally: the investor has weight $1/3$ and the two trustees together $2/3$. The separate shared-record tests score the queried player's action.

\textbf{One-shot and repeated decisions.} One-shot prompts omit history and future-round statements. Repeated actions are simultaneous. Following \citet{tewolde2026coopeval} and \citet{liu2026memorycurse}, prompts describe probabilistic continuation and a long-run payoff objective. Our prompts state a 99\% probability of another round; recording uses a fixed 200 rounds without announcing this endpoint. The history limit (HL) caps visible completed rounds at two (HL2) or forty (HL40); early in play, agents see all available rounds up to that cap. Each configuration uses five trajectory seeds or 20 one-shot draws. Section~\ref{sec:f4} defines the separate final-round and continuing probes.

\textbf{Response formats.} In the main one-shot and repeated-game evaluation, the model gives a rationale and then selects a legal action label. The shared-record tests use game-native JSON actions; the PGG shown-versus-hidden test also requests a rationale before the action. Sampling uses temperature 0.7, top-$p$ 0.8, and top-$k$ 20.

\textbf{What we measure.} The cooperation rate is the fraction of decisions taking the game's designated cooperative action. We report rates as percentages and changes as percentage points (pp): a fall from 70\% to 60\% is a change of $-10$ pp. The six binary-action games, including Trust, form our \emph{cooperation panel}; overall cooperation and history summaries weight these games equally. TD instead measures the chosen claim, normalized from zero for the lowest legal claim to one for the highest. Tables and figures multiply this value by 100 and report it separately.

\textbf{How to read the summaries.} Each lineage contributes one estimate to an overall mean. We first average runs and payoff settings within that lineage; summaries of the six binary games then weight those games equally. To compare stages, we compute their difference within each lineage before averaging the seven differences. Tables report the mean and, where shown, the SD across these seven estimates. The SD describes how much the models differ; a count such as 6/7 says how many lineages show the stated direction. Intervals give a range for the mean, based on either variation across lineages or repeated sampling with the same models, as specified in each table.

\textbf{Complementary stage comparisons.} Self-play measures the group behavior each checkpoint produces under common game rules and history limits. Every player uses the same checkpoint, so changing stages replaces both the focal player and its partners and changes the histories generated through interaction. The shared-record tests in Section~\ref{sec:f4} hold the presented past records fixed to compare checkpoints on common inputs.

\section{Cooperation Rates and Payoff Responses}
\label{sec:f1}
At HL40, Table~\ref{tab:surfaces} reports mean cooperation at four values of each game's main payoff parameter, labeled $D_1$--$D_4$; the parameter row gives the game-specific values. From $D_1$ to $D_4$, cooperation costs rise, benefits fall, or noncooperation becomes more rewarding. For example, lower PGG marginal per-capita return (MPCR)---each player's return per unit contributed---raises the private contribution cost.

\textbf{Cooperation levels.} Compare stages vertically within the same column. In every symmetric binary game, SFT and Aligned are below Base at all four displayed payoff values: they cooperate less under the same incentives. Averaging seven lineages at each complete binary-game configuration, SFT is below Base in 66/68 and Aligned in 64/68. Trust shows the same game-level ordering; TD instead has higher post-trained mean normalized claims.

\textbf{Responses to changing payoffs.} Read across a row, then compare its Drop with the other stages'. For each lineage and stage, a straight-line fit uses all four payoff values at their actual numerical spacing. The \emph{fitted drop} is the fitted rate at $D_1$ minus that at $D_4$, in pp. Positive values indicate a downward fitted trend; near-zero values indicate little net fitted change. Individual steps can still rise. \begin{table}[!htbp]
\centering
\caption{\textbf{Cooperation across four payoff values at HL40.} Entries are means across seven lineages (cooperation in \%; TD normalized claims multiplied by 100), after averaging seeds and secondary parameters. The final column is the fitted decline across all four points (pp); the numerical parameter spacing is retained.}\label{tab:surfaces}
\fontsize{8}{9.2}\selectfont
\setlength{\tabcolsep}{3pt}
\begin{minipage}[t]{0.49\linewidth}
\centering
\begin{tabular*}{\linewidth}{@{}l@{\extracolsep{\fill}}rrrrr@{}}\toprule
Game / stage & $D_1$ & $D_2$ & $D_3$ & $D_4$ & Drop \\
\midrule
\textbf{PGG}: MPCR & 0.8 & 0.6 & 0.5 & 0.4 & -- \\
\cmidrule{1-6}
Base (B) & $59.8$ & $58.6$ & $57.1$ & $57.2$ & $+2.9$ \\
SFT (S) & $43.2$ & $38.6$ & $35.4$ & $33.7$ & $+9.8$ \\
Aligned (A) & $42.2$ & $38.7$ & $35.6$ & $31.4$ & $+10.5$ \\
\midrule
\textbf{CPR}: $\theta$ & 0.2 & 0.4 & 0.6 & 0.8 & -- \\
\cmidrule{1-6}
Base (B) & $68.0$ & $63.7$ & $62.3$ & $65.6$ & $+2.6$ \\
SFT (S) & $31.4$ & $33.2$ & $30.5$ & $27.1$ & $+4.6$ \\
Aligned (A) & $23.6$ & $26.2$ & $24.6$ & $19.5$ & $+4.2$ \\
\midrule
\textbf{Donation-PD}: $R$ & 2.4 & 2 & 1.4 & 1.1 & -- \\
\cmidrule{1-6}
Base (B) & $54.3$ & $52.9$ & $53.7$ & $54.3$ & $-0.2$ \\
SFT (S) & $15.9$ & $12.6$ & $13.6$ & $13.4$ & $+1.8$ \\
Aligned (A) & $11.7$ & $8.9$ & $9.3$ & $8.7$ & $+2.3$ \\
\bottomrule\end{tabular*}
\end{minipage}\hfill
\begin{minipage}[t]{0.49\linewidth}
\centering
\begin{tabular*}{\linewidth}{@{}l@{\extracolsep{\fill}}rrrrr@{}}\toprule
Game / stage & $D_1$ & $D_2$ & $D_3$ & $D_4$ & Drop \\
\midrule
\textbf{Volunteer}: $\rho$ & 0.125 & 0.375 & 0.625 & 0.875 & -- \\
\cmidrule{1-6}
Base (B) & $80.6$ & $76.6$ & $75.1$ & $70.2$ & $+9.8$ \\
SFT (S) & $51.9$ & $49.3$ & $44.2$ & $42.3$ & $+10.2$ \\
Aligned (A) & $46.0$ & $42.6$ & $36.5$ & $33.8$ & $+12.8$ \\
\midrule
\textbf{Assurance}: $\lambda$ & 0.4 & 0.2 & 0.1 & 0.05 & -- \\
\cmidrule{1-6}
Base (B) & $87.2$ & $85.7$ & $86.8$ & $87.3$ & $+0.0$ \\
SFT (S) & $80.8$ & $78.3$ & $71.5$ & $75.0$ & $+7.7$ \\
Aligned (A) & $85.2$ & $80.0$ & $70.4$ & $76.4$ & $+12.0$ \\
\midrule
\textbf{TD}: $\tau$ & 0.025 & 0.05 & 0.1 & 0.2 & -- \\
\cmidrule{1-6}
Base (B) & $39.6$ & $39.8$ & $40.7$ & $37.9$ & $+1.7$ \\
SFT (S) & $67.2$ & $66.9$ & $67.0$ & $67.4$ & $-0.3$ \\
Aligned (A) & $62.6$ & $62.1$ & $61.4$ & $60.4$ & $+2.2$ \\
\bottomrule\end{tabular*}
\end{minipage}
\end{table}

In each symmetric binary game, both post-trained mean drops are positive and larger than Base's: cooperation falls more as per-round costs rise, benefits fall, or noncooperation becomes more rewarding (from $D_1$ to $D_4$). To see the stage gaps in Table~\ref{tab:surfaces}, subtract Base from SFT or Aligned within each column. In PGG, the SFT-minus-Base gap changes from $-16.6$ pp at $D_1$ to $-23.5$ pp at $D_4$: SFT falls further below Base. Comparing the Drop entries summarizes how these gaps change using all four points. Volunteer has a large cooperation gap but nearly parallel SFT and Base trends. Appendix~\ref{app:centered} reports these responses separately for each lineage.

\textbf{Do less cooperative groups also earn less?} PGG and Donation-PD let us answer this directly: at a fixed payoff setting, every contribution adds a fixed positive amount to the group's total, so the contribution rate determines average points earned per player. Other games require more information about joint actions; for example, Assurance payoffs depend on whether enough players contribute to complete the project. We therefore convert rates to returns in these two games before averaging settings. Both post-trained stages earn fewer points than Base on average at HL40, with decreases more consistent across lineages in Donation-PD.

\section{Post-Training Amplifies the Memory Curse}
\label{sec:f2}
Does the difference in cooperation between stages also change when more past interaction is visible? We compare separate 200-round HL2 and HL40 runs of the same checkpoints at the same complete payoff configurations. Following \citet{liu2026memorycurse}, the \emph{memory curse} means lower cooperation with a longer visible history. Averaging the six binary games equally within each lineage, mean cooperation falls by about 8 pp for SFT and 11 pp for Aligned, compared with about 1 pp for Base.

Both post-trained stages decline in all seven lineage summaries and decline more than Base in six; Base cooperation increases in five. Most of the average amplification is already present at SFT: its decline exceeds Base's by 6.93 pp, compared with a further 2.82 pp from SFT to Aligned. Table~\ref{tab:history-headline} shows that both post-trained stages' additional declines relative to Base have intervals below zero. The smaller difference between SFT and Aligned has an interval spanning zero. Appendix~\ref{app:model-first} gives the history changes for each lineage, including InternLM2's smaller post-trained declines.
\begin{table}[!htbp]
\centering\small
\setlength{\tabcolsep}{5pt}
\caption{\textbf{The memory curse and its amplification after post-training.} The first three rows report HL40 minus HL2; the last three compare these changes between stages. Entries give mean $\pm$ SD across seven lineages, in pp, after averaging the six binary games equally within each lineage. Stage differences are calculated within lineage. The exploratory 95\% $t$ intervals use these seven estimates; the last column counts negative estimates.}
\label{tab:history-headline}
\begin{tabular}{@{}lrrc@{}}\toprule
History change or paired difference & Mean $\pm$ SD & 95\% interval & Negative\\\midrule
Base: HL40 minus HL2 & $-1.17\pm3.92$ & $[-4.79,+2.45]$ & 2/7\\
SFT: HL40 minus HL2 & $-8.11\pm5.46$ & $[-13.16,-3.05]$ & 7/7\\
Aligned: HL40 minus HL2 & $-10.93\pm6.40$ & $[-16.85,-5.01]$ & 7/7\\\midrule
SFT minus Base history change & $-6.93\pm7.30$ & $[-13.68,-0.19]$ & 6/7\\
Aligned minus Base history change & $-9.76\pm7.47$ & $[-16.67,-2.85]$ & 6/7\\
Aligned minus SFT history change & $-2.82\pm4.37$ & $[-6.87,+1.22]$ & 6/7\\\bottomrule
\end{tabular}
\end{table}

Let $p_s(L)$ be the expected cooperation rate, as a fraction, at stage $s$ with history limit $L$. The history change $H_s$ compares HL40 with HL2; $K_{s/t}$ compares that change between stages:
\begin{equation}
H_s=p_s(40)-p_s(2),\qquad K_{s/t}=H_s-H_t. \label{eq:history}
\end{equation}
A negative $H_s$ means that showing more history lowers cooperation at that stage---the memory curse. $K_{S/B}$ compares SFT with Base, $K_{A/B}$ compares Aligned with Base, and $K_{A/S}$ compares Aligned with SFT. We multiply these rate differences by 100 when reporting pp. For example, $K_{S/B}=-0.10$, displayed as $-10$ pp, means that showing more history reduces SFT cooperation by ten percentage points more than Base cooperation.

\begin{figure}[!htbp]
\centering
\includegraphics[width=\linewidth]{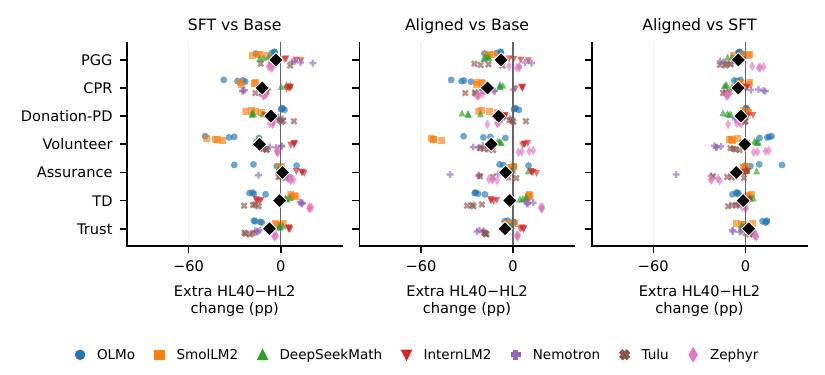}
\caption{\textbf{Which stage loses more cooperation when shown a longer history?} Panels compare SFT vs Base ($K_{S/B}$), Aligned vs Base ($K_{A/B}$), and Aligned vs SFT ($K_{A/S}$). Left of zero means a more negative change from HL2 to HL40 at the first-named stage: a larger decrease or smaller increase. TD reports normalized claims. Each game/panel has 28 colored points: seven lineages $\times$ four primary payoff values, averaging five seeds and secondary parameters. Black diamonds average points equally within and across lineages. All three panels use the same percentage-point scale.}
\label{fig:history}\label{fig:history-primary}
\end{figure}

Figure~\ref{fig:history} separates the games and payoff values. A point left of zero means that cooperation falls more, or rises less, at the first-named stage than at the second. Black diamonds show game averages; colored points show individual lineage and payoff results. SFT's mean history change is more negative than Base's in PGG, CPR, Donation-PD, and Volunteer; Assurance has the opposite ordering. Aligned declines more than Base and SFT in all five symmetric binary-game means; in Trust, SFT declines more than Aligned. Longer history also lowers post-trained models' mean returns in PGG and Donation-PD. 

\textbf{A possible feedback mechanism.} A model that reduces cooperation after a partner defects changes what that partner sees next. If neither readily resumes cooperating, their responses can reinforce one another over time. Post-training may change this balance between withdrawal and recovery. Human PGG experiments show a related pattern: contributing less than others helps drive declining cooperation \citep{fischbacher2010social}.

\textbf{Beyond three checkpoints.} Appendix~\ref{app:cascade} follows Nemotron-Cascade through seven training stages. Averaged across the six binary games, all six post-Base checkpoints show a larger mean HL2-to-HL40 decline than Base. Yet the instruction-following RL stage after RLHF raises mean cooperation in one-shot, HL2, and HL40 evaluations. Later training can therefore restore cooperation levels while a stronger history-related decline remains. Figure~\ref{fig:cascade-common} shows how levels, payoff responses, and history effects change along this training path.

\section{Cooperation across Interaction Settings}
\label{sec:context}

The history result raises a further question: does a one-shot cooperation advantage persist in actual interaction? We focus on Aligned versus SFT to examine training after SFT, then compare their choices given the same past records.

\subsection{Aligned versus SFT in one-shot and repeated play}
\label{sec:f3}

Table~\ref{tab:f3-allseat} asks whether Aligned cooperates more than SFT. Positive entries mean that it does. The first two columns give this difference in one-shot decisions and repeated play; the third shows how much it changes. We calculate changes within each lineage before averaging them. A negative change means that moving to repeated play shifts the difference toward SFT.

The difference decreases in all five symmetric binary-game means. In PGG, CPR, and Volunteer, Aligned cooperates more than SFT in one-shot decisions but less in repeated play. In Donation-PD, it cooperates less in both; Assurance changes little. Volunteer has the clearest pattern: the difference falls by about 10 pp, with all seven lineages showing a decrease. Lineage directions vary in the other games. TD's mean claim difference also decreases, whereas Trust's mean intention difference increases. A one-shot cooperation advantage can thus disappear when checkpoints actually interact.

\begin{table}[!htbp]
\centering
\small
\setlength{\tabcolsep}{5pt}
\caption{\textbf{Does Aligned cooperate more than SFT in one-shot decisions and repeated play?} The first two columns report Aligned minus SFT; positive values mean more cooperation at Aligned. Change is repeated minus one-shot, computed within each lineage before averaging; negative values mean a smaller Aligned-minus-SFT difference in repeated play. These columns give mean $\pm$ SD across seven model lineages, in pp. SD describes variation in effect magnitude and direction across model lineages; the last column counts lineages with a negative change. TD measures normalized claims; Trust averages roles.}
\label{tab:f3-allseat}
\begin{tabular}{@{}lrrrc@{}}
\toprule
Game & One-shot & Repeated (HL40) & Change & Lineages with\\
 & Aligned $-$ SFT & Aligned $-$ SFT & Repeated $-$ one-shot & a decrease\\
\midrule
PGG & $+8.62\pm6.78$ & $-0.77\pm20.44$ & $-9.39\pm20.00$ & 5/7 \\
CPR & $+3.47\pm22.35$ & $-7.08\pm14.96$ & $-10.55\pm19.14$ & 5/7 \\
Donation-PD & $-1.71\pm7.11$ & $-4.21\pm5.51$ & $-2.51\pm9.25$ & 4/7 \\
Volunteer & $+2.66\pm14.47$ & $-7.25\pm11.63$ & $-9.91\pm5.18$ & 7/7 \\
Assurance & $+1.93\pm13.67$ & $+1.59\pm8.14$ & $-0.35\pm8.96$ & 3/7 \\
TD & $+2.48\pm9.71$ & $-5.50\pm8.82$ & $-7.98\pm12.26$ & 5/7 \\
Trust & $-3.91\pm8.47$ & $+0.21\pm12.99$ & $+4.12\pm9.03$ & 2/7 \\
\bottomrule
\end{tabular}
\end{table}

\textbf{Interpretation.} Repeated play supplies feedback from partners: a model's choices and its partners' responses become part of the history used for later decisions. Training recipes also differ across lineages. SFT demonstrations already teach desired behavior, and later training uses different preferences and rewards. These differences may help explain why the Aligned--SFT comparison varies across models.

\subsection{Instructions about future rounds have game-specific effects}
\label{sec:f4}
\subsubsection{Final-round and continuing decisions}

Self-play gives different checkpoints different histories. We now supply the same scripted records and ask for one current decision under different future instructions across all seven games. \textsc{Final} says that the interaction ends after the current round. \textsc{Continue} says that there is a 99\% probability of another round, that actions will be visible in future rounds, and that the goal is to maximize long-run points. We keep the payoff, checkpoint, scripted past records, and current round number (41) the same between the two prompts. The comparison measures the effect of changing these future-interaction instructions together.

We supply either no past records or the last 2, 10, or 40 rounds of seven scripted behavior patterns.\footnote{The patterns are: everyone cooperates; no one cooperates; only the focal player cooperates; only the partners cooperate; odd- and even-numbered seats take opposite actions and alternate each round; everyone cooperates for 20 rounds, then only the focal player cooperates; and no one cooperates for 20 rounds, then everyone cooperates. TD uses highest and lowest claims for the two actions.} This gives $1+3\times7=22$ equally weighted conditions, with 20 action draws per prompt. This probe uses PGG MPCR values $\{0.4,0.5,0.6,0.7,0.8\}$ and CPR $\theta\in\{0.2,0.5,0.8\}$; other payoff parameters follow Appendix~\ref{app:games}. Table~\ref{tab:continuation} averages these single-decision tests; the earlier repeated-game results measure 200 rounds of actual interaction. \begin{table}[!htbp]
\centering\footnotesize
\setlength{\tabcolsep}{3pt}
\caption{\textbf{How instructions about future rounds change the current decision.} \textsc{Final} says this is the last round; \textsc{Continue} states a 99\% chance of another. Cooperation is in percent; TD claims are rescaled from 0 (lowest allowed) to 100 (highest). $\Delta$ is \textsc{Continue} minus \textsc{Final}, in pp. Entries are mean $\pm$ SD across seven model lineages, after averaging payoff settings and 22 history conditions within lineage. SD summarizes variation across model lineages. The first three numeric columns also average stages; the last three keep stages separate. Trust equally weights the investor and a representative trustee.}\label{tab:continuation}
\begin{tabular}{lrrrrrr}\toprule
Game & \textsc{Final} & \textsc{Continue} & Mean $\Delta$ & $\Delta$ Base & $\Delta$ SFT & $\Delta$ Aligned \\
\midrule
PGG & $73.7\pm14.3$ & $73.2\pm14.4$ & $-0.5\pm4.0$ & $-0.1\pm3.4$ & $+2.5\pm6.4$ & $-3.9\pm6.8$ \\
CPR & $43.9\pm18.4$ & $59.0\pm15.2$ & $+15.1\pm9.7$ & $+6.2\pm4.0$ & $+14.0\pm10.4$ & $+25.1\pm20.6$ \\
Donation-PD & $61.0\pm21.8$ & $64.0\pm14.5$ & $+3.0\pm9.9$ & $-0.8\pm7.2$ & $+5.5\pm12.7$ & $+4.4\pm12.2$ \\
Volunteer & $78.3\pm13.7$ & $81.3\pm10.5$ & $+3.0\pm7.5$ & $+3.6\pm5.1$ & $+2.8\pm10.8$ & $+2.8\pm11.0$ \\
Assurance & $72.9\pm18.0$ & $75.1\pm12.2$ & $+2.2\pm10.8$ & $+1.0\pm6.3$ & $+3.8\pm16.6$ & $+1.6\pm12.9$ \\
TD & $53.2\pm14.6$ & $40.7\pm9.8$ & $-12.6\pm8.2$ & $-8.4\pm9.5$ & $-11.4\pm7.2$ & $-17.9\pm14.3$ \\
Trust & $73.9\pm11.9$ & $75.4\pm9.5$ & $+1.5\pm5.9$ & $-0.2\pm1.8$ & $+3.2\pm9.3$ & $+1.5\pm8.0$ \\
\bottomrule\end{tabular}
\end{table}

In Table~\ref{tab:continuation}, the effect of continuing instructions is the \textsc{Continue} rate minus the \textsc{Final} rate, denoted $\Delta$ and reported in pp. Continuing instructions increase CPR restraint by about 15 pp on average and lower TD claims by about 13 pp. Both directions hold at each of the three stages after averaging history conditions. In CPR, the increase is larger at SFT than Base and larger again at Aligned. In PGG, the small average effect combines an increase at SFT with a decrease at Aligned. Thus the same instructions about future interaction can raise or lower cooperation, depending on the game and checkpoint.

Human repeated-PD experiments find higher cooperation at greater actual continuation probabilities after participants gain experience \citep{dalbo2011evolution}. Our test asks for one choice after telling the model either that this is the final round or that there is a 99\% chance of another round. These continuing instructions raise CPR cooperation, lower TD claims, and change PGG cooperation little on average (Table~\ref{tab:continuation}).

\subsubsection{Past records and instructions about future interaction}
\label{sec:visibility}

Does showing past records have a different effect under final-round and continuing instructions? We give SFT and Aligned fresh PGG decisions at $(n,r)=(3,1.4),(3,2.2),(5,1.8),(5,4.2)$ under four combinations: records shown or hidden, each with \textsc{Final} or \textsc{Continue}. Records come from separately collected SFT and Aligned games at these settings. For each lineage and payoff, we sample eight records from each source stage in each time bin: rounds 1--20, 21--50, 51--100, and 101--200. Each checkpoint generates 32 answers per record under each of the four conditions. Payoffs and source stages receive equal weight; time bins are weighted by their lengths. \textsc{Real} shows saved actions, earned points, and summaries; \textsc{Masked} removes them but keeps rules, role, round number, and future instructions. Removing records also shortens the prompt. Table~\ref{tab:visibility-components} subtracts hidden-record from shown-record cooperation: negative means less cooperation when records are shown. \begin{table}[!htbp]
\centering\small
\setlength{\tabcolsep}{3pt}
\caption{\textbf{How showing past records changes PGG cooperation.} \textsc{Real} shows the records; \textsc{Masked} hides them. The first two numeric columns give rates (\%); the last subtracts the second from the first to give the effect of showing records (pp). Negative values mean less cooperation with records visible. Entries are mean $\pm$ SD across seven lineages. All four conditions use the same payoff and record weights.}\label{tab:visibility-components}
\begin{tabular}{llrrr}\toprule
Future instruction & Stage & \textsc{Real} & \textsc{Masked} & \textsc{Real}$-$\textsc{Masked} \\
\midrule
\textsc{Continue} & SFT (S) & $41.9\pm31.7$ & $58.2\pm27.6$ & $-16.3\pm12.3$ \\
 & Aligned (A) & $41.4\pm35.3$ & $63.3\pm29.9$ & $-22.0\pm12.9$ \\
\midrule
\textsc{Final} & SFT (S) & $35.8\pm22.2$ & $51.5\pm22.0$ & $-15.7\pm21.9$ \\
 & Aligned (A) & $34.6\pm26.4$ & $46.6\pm23.8$ & $-12.0\pm32.0$ \\
\bottomrule\end{tabular}
\end{table}

Showing past records lowers mean contribution at both stages. Under \textsc{Continue}, the mean reduction is larger for Aligned than for SFT; under \textsc{Final}, it is larger for SFT. The ordering of these average history effects therefore reverses when the same decision is described as final rather than potentially continuing.

Subtracting SFT's record effect from Aligned's gives one difference for each frame. This difference falls by 9.42 pp on average when \textsc{Final} becomes \textsc{Continue}, with the same direction in five of seven lineages. The 97.5\% interval from resampling lineages is $[-24.29,+2.45]$ pp and includes zero.

\textbf{Why the final-round result is informative.} In final-round PGG, withholding one's contribution always earns $1-r/n>0$ more points this round, holding the others' actions fixed. Past records still change the models' choices. This pattern is consistent with responses to past behavior persisting under explicit final-round instructions.

\section{Related Work}
\label{sec:related}

\textbf{Cooperation and memory in language agents.} Prior studies examine how language agents respond to social objectives, partners, and repeated interaction \citep{phelps2023machine,fontana2024pd,akata2024repeated,piatti2024cooperate}. A fuller comparison appears in Appendix~\ref{app:detailed-related}. This literature evaluates both cooperation-sustaining mechanisms, such as repetition, reputation, mediation, and contracting in CoopEval \citep{tewolde2026coopeval}, and history effects, including the memory curse \citep{liu2026memorycurse}. CoopShift asks whether the cooperation change from showing more history differs across Base, SFT, and Aligned checkpoints within a released lineage. We complement this comparison with probes that give SFT and Aligned the same past records and with the payoffs earned during self-play. These measure, respectively, choices under common historical inputs and outcomes generated by interacting groups.

\textbf{Post-training and human behavior.} Instruction and preference optimization train models to produce desired responses \citep{ouyang2022training,rafailov2023dpo}. We ask how released checkpoints behave when their decisions affect other players. Human experiments provide comparisons for specific responses to costs, benefits, and past interaction.

\section{Discussion and Conclusion}
\label{sec:outcomes}
Differences in cooperation between checkpoints depend on both costs and benefits and the history agents can see. Most of the average memory-curse amplification appears between Base and SFT, making demonstration training central to understanding it.

\textbf{Interpreting Base behavior.} Base follows the same procedure of giving a rationale and then choosing an action. Its small average history effect combines different responses: InternLM2 Base and Nemotron's Qwen3 Base decline by about 7.6 and 6.0 pp. Some Base checkpoints therefore respond to history.

\textbf{Conclusion.} Post-training amplifies the memory curse in our six-binary-game panel. At HL40, both post-trained stages cooperate less than Base on average in each binary game. In the symmetric binary games, their fitted cooperation rates also fall more than Base's, on average, as per-round cooperation costs rise, benefits fall, or noncooperation becomes more rewarding. Evaluation should track how cooperation changes with payoffs and visible history and what returns it produces.
\section*{Acknowledgments}
Jiayuan Liu, Emanuel Tewolde, and Vincent Conitzer thank the Cooperative AI Foundation, Macroscopic Ventures (formerly Polaris Ventures / the Center for Emerging Risk Research) and Jaan Tallinn's donor-advised fund at Founders Pledge for financial support.
Shiyi Du is partially supported by the SoftBank Group--Arm Fellowship. Tianqin Li is supported by NSF CISE RI 1816568 and NIH R01 EY030226–01A1.

\label{main-text-end}
\section*{Reproducibility Statement}
Section~\ref{sec:design} describes the evaluation conditions, response formats, sampling settings, and aggregation rules. Appendices~\ref{app:models} and~\ref{app:games} identify the checkpoints and specify the game payoffs. Appendix~\ref{app:model-first} gives history changes by lineage; Table~\ref{tab:surfaces} lists the payoff values used for the response fits, and Appendix~\ref{app:centered} gives payoff responses by lineage. Appendix~\ref{app:cascade} specifies the seven-stage Nemotron-Cascade extension and its common payoff grid. Section~\ref{sec:f4} describes the final-round, continuing, and shown-versus-hidden record comparisons.

\section*{AI Use Statement}
Generative AI assisted conceptual framing, hypothesis refinement, experimental-design feedback, elementary payoff derivations and repeated-game reasoning, implementation, data cleaning and reformatting, statistical cross-checks, result interpretation, literature synthesis, manuscript drafting, and figure/table preparation. The evaluated agents themselves generate the synthetic game decisions. The authors are responsible for scientific validation and the paper's text, figures, tables, and claims.

\bibliographystyle{coopshift}
\bibliography{references,interpretation_references,human_context_references,editorial_references}
\clearpage
\appendix
\section{Detailed Related Work}
\label{app:detailed-related}
\label{app:human-boundaries}
\subsection{Post-training and social-agent evaluation}
Instruction tuning and learning from human feedback improve evaluated instruction following and preferences \citep{ouyang2022training}; DPO optimizes preferences without the same explicit reward-model-and-RL pipeline \citep{rafailov2023dpo}. Generalization comparisons of SFT and RL \citep{chu2025sft}, and the distinction between high-probability reasoning success and solutions recoverable by sampling \citep{yue2025reasoning}, emphasize defining the evaluation target. We compare released stages on their own generated cooperation, payoff responses, and history responses across several preference- and reward-optimization objectives.

Economic-game evaluations already demonstrate sensitivity to objectives, prompts, partners and repetition \citep{phelps2023machine,fontana2024pd,akata2024repeated}. SOTOPIA evaluates interactive goals and relationships \citep{zhou2024sotopia}; GovSim adds resource sustainability over time \citep{piatti2024cooperate}, unlike our stationary CPR payoff. Social training studies investigate behavior cloning, self-reinforcement, reward design, and incentive-supported coordination \citep{wang2024sotopiapi,yu2025sotopiarL,liang2025everyone}. CoopEval compares repetition, reputation, mediation and contracting \citep{tewolde2026coopeval}. We complement these studies by comparing released training stages under common game rules and evaluation procedures.

\subsection{Memory, conditional cooperation, and recovery}
\citet{liu2026memorycurse} introduce the memory curse as lower cooperation with more visible interaction history and study historical content, reasoning, and interventions. We ask how this decline changes along a training sequence. Our comparisons $K_{S/B}$ and $K_{A/B}$ measure whether longer histories reduce cooperation more after post-training than at Base; $K_{A/S}$ compares SFT with its later Aligned checkpoint. We build on part of that experimental framework with newly collected game data. Studies of reasoning in public goods and norm enforcement further motivate examining both the rationale and the selected action \citep{guzman2025corrupted,li2025giving}.

Human conditional-contribution schedules \citep{fischbacher2001conditional}, separately elicited preferences and beliefs \citep{fischbacher2010social}, and distributional/reciprocity choices \citep{charness2002social} motivate separating an opening action from its subsequent feedback. \citet{ma2021limited} display zero to four encounters within anonymous eight-person groups and find an intermediate-history peak; we instead compare caps of two and forty in fixed groups. Noisy repeated-PD experiments motivate studying tolerance and recovery after temporary defection \citep{fudenberg2012slow}. Our generated self-play and fixed-record probes resolve different parts of that question: the former changes all players and their resulting histories; the latter compares responses to a shared input.

\subsection{Game-specific payoff responses and human comparisons}
Human public-goods studies separately manipulate MPCR/group size in ten-period play \citep{isaac1988group} and private/external returns in ten no-feedback one-shot decisions \citep{goeree2002private}; a single-decision study reports no statistically detected MPCR response \citep{struwe2024increasing}. These designs motivate identifying both the economic margin and the interaction protocol. Sequential dilemmas vary a second mover's private deviation gain and harm to a cooperative first mover \citep{baader2026conditional}. Their positive mutual-cooperation reference supports relative-gain measures; our response statistic instead uses the tested span of each game's raw parameter.

Human TD claims fall with larger reward/penalty sizes under random rematching \citep{capra1999anomalous}; related LLM signatures are studied by \citet{roberts2025strategic}. Human Trust experiments elicit investment and returns in a sequential two-person design \citep{berg1995trust}; ours elicits one investor's and two trustees' simultaneous contingent intentions. Behavioral-test similarity \citep{mei2024turing} and prediction of observed human choices \citep{shapira2026normative,binz2026posttraining} are different evaluation targets; the latter also includes one-shot/first-round settings favoring aligned models. Our comparisons concern specific payoff and history responses under each study's stated protocol.

\subsection{Continuation, presentation, and shared-input comparisons}
Repeated-game theory distinguishes equilibrium support from equilibrium selection \citep{fudenberg1986folk}. Human repeated-PD evidence studies this distinction empirically \citep{dalbo2011evolution}. Public-goods studies use different finite or probabilistic stopping procedures and report different cooperation patterns \citep{lugovskyy2017experimental,anderson2024cooperation}. They motivate specifying both what agents are told about future rounds and how collection actually stops. Our canonical data record 200 rounds under continuing instructions; the shared-record tests vary a current decision's instructions at fixed presented history and payoff.

Externality-framing experiments change how a decision is presented while keeping its potential payoffs fixed \citep{andreoni1995warm}. Our PGG comparison shows or hides past records, which also changes prompt length. In the final-round condition, contributing still changes the player's current payoff by $r/n-1$, regardless of the records shown. These tests ask how a model responds to the information it receives, alongside our measurements of cooperation during actual interaction.

\section{Model Checkpoints and Training Ancestry}
\label{app:models}
Table~\ref{tab:modelcatalog} identifies the 21 released checkpoints in the primary panel: one Base, one SFT, and one Aligned checkpoint per lineage. A colon separates a Hugging Face repository identifier from a checkpoint subfolder. In particular, OLMo-3-7B Base denotes \path{allenai/Olmo-3-1025-7B}; \texttt{1025} is part of its release identifier. The table also names each Aligned training objective.
\begingroup\small
\newcommand{\checkpointrepo}[1]{\href{https://huggingface.co/#1}{\nolinkurl{#1}}}
\makeatletter
\let\modelcatalogcaption\LT@makecaption
\def\LT@makecaption#1#2#3{\modelcatalogcaption{#1}{\normalsize #2}{\normalsize #3}}
\makeatother
\setlength{\tabcolsep}{3pt}
\setlength{\LTcapwidth}{\linewidth}
\begin{longtable}{p{0.20\linewidth}p{0.19\linewidth}p{0.55\linewidth}}
\caption{All 21 checkpoints in the primary panel: one Base, SFT, and Aligned release per lineage. Identifiers link to the Hugging Face repositories; Nemotron's SFT and RLHF suffixes identify checkpoint subdirectories. Aligned training objectives appear in parentheses.}\label{tab:modelcatalog}\\
\toprule
Lineage & Stage & Hugging Face repository \\
\midrule
\endfirsthead
\toprule
Lineage & Stage & Hugging Face repository \\
\midrule
\endhead
\bottomrule
\endfoot
OLMo-3-7B & Base & \checkpointrepo{allenai/Olmo-3-1025-7B} \\
OLMo-3-7B & SFT & \checkpointrepo{allenai/Olmo-3-7B-Instruct-SFT} \\
OLMo-3-7B & Aligned (DPO) & \checkpointrepo{allenai/Olmo-3-7B-Instruct-DPO} \\
\addlinespace[2pt]
SmolLM2-1.7B & Base & \checkpointrepo{HuggingFaceTB/SmolLM2-1.7B} \\
SmolLM2-1.7B & SFT & \checkpointrepo{HuggingFaceTB/SmolLM2-1.7B-sft-only} \\
SmolLM2-1.7B & Aligned (DPO) & \checkpointrepo{HuggingFaceTB/SmolLM2-1.7B-Instruct} \\
\addlinespace[2pt]
Mistral/Zephyr-7B & Base & \checkpointrepo{mistralai/Mistral-7B-v0.1} \\
Mistral/Zephyr-7B & SFT & \checkpointrepo{HuggingFaceH4/mistral-7b-sft-beta} \\
Mistral/Zephyr-7B & Aligned (DPO) & \checkpointrepo{HuggingFaceH4/zephyr-7b-beta} \\
\addlinespace[2pt]
Llama/T\"ulu-3-8B & Base & \checkpointrepo{meta-llama/Llama-3.1-8B} \\
Llama/T\"ulu-3-8B & SFT & \checkpointrepo{allenai/Llama-3.1-Tulu-3-8B-SFT} \\
Llama/T\"ulu-3-8B & Aligned (DPO) & \checkpointrepo{allenai/Llama-3.1-Tulu-3-8B-DPO} \\
\addlinespace[2pt]
InternLM2-7B & Base & \checkpointrepo{internlm/internlm2-base-7b} \\
InternLM2-7B & SFT & \checkpointrepo{internlm/internlm2-chat-7b-sft} \\
InternLM2-7B & Aligned (online-RLHF) & \checkpointrepo{internlm/internlm2-chat-7b} \\
\addlinespace[2pt]
DeepSeekMath-7B & Base & \checkpointrepo{deepseek-ai/deepseek-math-7b-base} \\
DeepSeekMath-7B & SFT & \checkpointrepo{deepseek-ai/deepseek-math-7b-instruct} \\
DeepSeekMath-7B & Aligned (GRPO/RL) & \checkpointrepo{deepseek-ai/deepseek-math-7b-rl} \\
\addlinespace[2pt]
Nemotron-Cascade-8B & Base & \checkpointrepo{Qwen/Qwen3-8B-Base} \\
Nemotron-Cascade-8B & SFT & \href{https://huggingface.co/nvidia/Nemotron-Cascade-8B-Intermediate-ckpts/tree/main/SFT}{\nolinkurl{nvidia/Nemotron-Cascade-8B-Intermediate-ckpts:SFT}} \\
Nemotron-Cascade-8B & Aligned (RLHF) & \href{https://huggingface.co/nvidia/Nemotron-Cascade-8B-Intermediate-ckpts/tree/main/RLHF}{\nolinkurl{nvidia/Nemotron-Cascade-8B-Intermediate-ckpts:RLHF}} \\
\end{longtable}
\endgroup

\normalsize

\paragraph{Training stages across lineages.}
OLMo-3, SmolLM2, Zephyr, and T\"ulu-3 use DPO releases as their Aligned checkpoints. InternLM2 uses its online-RLHF Chat release, DeepSeekMath its GRPO-trained RL release, and Nemotron-Cascade the \texttt{RLHF} subfolder of its intermediate-checkpoint repository. OLMo-3 and T\"ulu-3 contribute their DPO checkpoints to this panel. SFT and Aligned identify released training stages with lineage-specific demonstrations and subsequent optimization objectives.

\paragraph{Ancestry details.}
The InternLM2 SFT model card documents training from InternLM2-Base, so we use \path{internlm/internlm2-base-7b}. Additional pretraining distinguishes the separately released \path{internlm/internlm2-7b} from this SFT parent. The SmolLM2 SFT release names an 8k-context parent, whereas the evaluated Base is the published 1.7B endpoint listed in Table~\ref{tab:modelcatalog}.
\FloatBarrier

\section{Game Payoffs and Parameter Grids}
\label{app:games}
Here $u_i$ is player $i$'s one-round payoff in game points. In binary-action games, $C$ denotes the designated cooperative action ($a_i=1$) and $D$ the alternative ($a_i=0$); these are explanatory labels for the implemented actions. We write $\mathbf{1}\{E\}$ for the indicator of event $E$: it equals one when the event holds and zero otherwise. Parameters are local to each game. Every numerical example below uses a configuration in the amended main grid.

\paragraph{Reading the matrices.}
For the first five games, rows give a focal player's action and columns give $m_{-i}=\sum_{j\ne i}a_j$, the number of other players choosing $C$. Each entry gives the focal player's payoff conditional on this count. In these symmetric games, this count summarizes the other players' actions, so the compact matrices specify every player's payoff for every joint action in the multiplayer game. TD entries instead give the ordered pair $(u_1,u_2)$, and Trust entries give $(u_1,u_2,u_3)$. The entries are deterministic one-round payoffs.

\subsection{Public-goods game (PGG)}
\label{app:payoffs-pgg}
Each of $n$ players contributes its unit endowment ($C$) or retains it ($D$). For $k=\sum_j a_j$ contributors and MPCR $\alpha=r/n$, the payoff is $u_i=1-a_i+\alpha k$. We cross $n\in\{3,5\}$ with $\alpha\in\{0.4,0.5,0.6,0.8\}$ in the amended main panel. A contribution increases the return to every player by $\alpha$, while its contributor gives up the unit endowment; its net private cost is $1-\alpha$.

\begin{center}
\begin{minipage}{\linewidth}
\centering\small
\textbf{General payoff matrix: focal payoff $u_i$.}\par\smallskip
\begin{tabular}{@{}lcccc@{}}
\toprule
Own action & $m_{-i}=0$ & $m_{-i}=1$ & $\cdots$ & $m_{-i}=n-1$ \\
\midrule
$C$: contribute & $\alpha$ & $2\alpha$ & $\cdots$ & $n\alpha$ \\
$D$: retain & $1$ & $1+\alpha$ & $\cdots$ & $1+(n-1)\alpha$ \\
\bottomrule
\end{tabular}
\par\medskip
\textbf{Example: $n=3$, $\alpha=0.6$, hence $r=1.8$.}\par\smallskip
\begin{tabular}{@{}lccc@{}}
\toprule
Own action & $m_{-i}=0$ & $m_{-i}=1$ & $m_{-i}=2$ \\
\midrule
$C$: contribute & $0.6$ & $1.2$ & $1.8$ \\
$D$: retain & $1.0$ & $1.6$ & $2.2$ \\
\bottomrule
\end{tabular}
\end{minipage}
\end{center}
For example, one contributor and two noncontributors receive $(0.6,1.6,1.6)$ when the contributor is Player 1.

\subsection{Common-pool resource game (CPR)}
\label{app:payoffs-cpr}
Three players choose restrained extraction $x_i=1$ ($C$) or high extraction $x_i=h$ ($D$), with $h\in\{1.25,1.5,2.5,3\}$. Their payoff is $u_i=x_i(V-\sum_jx_j)$, where $V=3h+1+2\theta$ and the amended main grid is $\theta\in\{0.2,0.4,0.6,0.8\}$. Thus $m_{-i}$ counts the other restrained extractors. The game uses a stationary resource-return rule applied separately in each round, with payoffs determined by that round's joint extraction.

\begin{center}
\begin{minipage}{\linewidth}
\centering\small
\textbf{General payoff matrix: focal payoff $u_i$.}\par\smallskip
\begin{tabular}{@{}lccc@{}}
\toprule
Own action & $m_{-i}=0$ & $m_{-i}=1$ & $m_{-i}=2$ \\
\midrule
$C$: extract $1$ & $V-1-2h$ & $V-2-h$ & $V-3$ \\
$D$: extract $h$ & $h(V-3h)$ & $h(V-1-2h)$ & $h(V-2-h)$ \\
\bottomrule
\end{tabular}
\par\medskip
\textbf{Example: $h=1.5$, $\theta=0.4$, hence $V=6.3$.}\par\smallskip
\begin{tabular}{@{}lccc@{}}
\toprule
Own action & $m_{-i}=0$ & $m_{-i}=1$ & $m_{-i}=2$ \\
\midrule
$C$: extract $1$ & $2.30$ & $2.80$ & $3.30$ \\
$D$: extract $1.5$ & $2.70$ & $3.45$ & $4.20$ \\
\bottomrule
\end{tabular}
\end{minipage}
\end{center}
In the example, all restrained extractors earn $3.30$ each, while all high extractors earn $2.70$ each. Holding the other two actions fixed, high extraction nevertheless gives the larger private payoff in every column.

\subsection{Donation Prisoner's Dilemma (Donation-PD)}
\label{app:payoffs-donation}
Each of three players starts with 2 points. Cooperating pays cost $c$ to give benefit $b$ to each other player; defecting pays no cost and gives nothing. The payoff is $u_i=2-ca_i+b\sum_{j\ne i}a_j$. We cross $c\in\{0.2,0.6,1.2\}$ with $R=2b/c\in\{1.1,1.4,2.0,2.4\}$. At fixed $c$, changing $R$ changes others' contribution returns while the immediate private difference remains $u_D-u_C=c$. Its oriented response thus measures sensitivity to external returns.

\begin{center}
\begin{minipage}{\linewidth}
\centering\small
\textbf{General payoff matrix: focal payoff $u_i$.}\par\smallskip
\begin{tabular}{@{}lccc@{}}
\toprule
Own action & $m_{-i}=0$ & $m_{-i}=1$ & $m_{-i}=2$ \\
\midrule
$C$: donate & $2-c$ & $2-c+b$ & $2-c+2b$ \\
$D$: do not donate & $2$ & $2+b$ & $2+2b$ \\
\bottomrule
\end{tabular}
\par\medskip
\textbf{Example: $c=0.6$, $R=2$, hence $b=0.6$.}\par\smallskip
\begin{tabular}{@{}lccc@{}}
\toprule
Own action & $m_{-i}=0$ & $m_{-i}=1$ & $m_{-i}=2$ \\
\midrule
$C$: donate & $1.4$ & $2.0$ & $2.6$ \\
$D$: do not donate & $2.0$ & $2.6$ & $3.2$ \\
\bottomrule
\end{tabular}
\end{minipage}
\end{center}

\subsection{Volunteer's Dilemma (Volunteer)}
\label{app:payoffs-volunteer}
Three players choose whether to volunteer. If at least one volunteers, \emph{every} player receives benefit $B$, and \emph{each} volunteer separately pays cost $c$. If nobody volunteers, all receive zero. Thus $u_i=B\mathbf1\{\sum_j a_j\ge1\}-ca_i$, with $B\in\{2,4,8\}$ and $\rho=c/B\in\{0.125,0.375,0.625,0.875\}$.

A volunteer always earns the net payoff $B-c=B(1-\rho)$, whether or not anyone else volunteers. A nonvolunteer earns $B$ if another player volunteers and zero otherwise. Each player receives the full benefit $B$ when provision occurs, and every volunteer pays its own full cost $c$. Across the four tested cost ratios, a volunteer therefore receives $0.875B$, $0.625B$, $0.375B$, or $0.125B$, respectively.

\begin{center}
\begin{minipage}{\linewidth}
\centering\small
\textbf{General payoff matrix: focal payoff $u_i$.}\par\smallskip
\begin{tabular}{@{}lccc@{}}
\toprule
Own action & $m_{-i}=0$ & $m_{-i}=1$ & $m_{-i}=2$ \\
\midrule
$C$: volunteer & $B-c$ & $B-c$ & $B-c$ \\
$D$: do not volunteer & $0$ & $B$ & $B$ \\
\bottomrule
\end{tabular}
\par\medskip
\textbf{Example: $B=4$, $\rho=0.375$, hence $c=1.5$.}\par\smallskip
\begin{tabular}{@{}lccc@{}}
\toprule
Own action & $m_{-i}=0$ & $m_{-i}=1$ & $m_{-i}=2$ \\
\midrule
$C$: volunteer & $2.5$ & $2.5$ & $2.5$ \\
$D$: do not volunteer & $0$ & $4$ & $4$ \\
\bottomrule
\end{tabular}
\end{minipage}
\end{center}
For example, $(C,D,D)$ gives $(2.5,4,4)$, $(C,C,D)$ gives $(2.5,2.5,4)$, and $(D,D,D)$ gives $(0,0,0)$. Volunteering is privately beneficial when nobody else volunteers; otherwise, not volunteering avoids the cost.

Volunteer is a threshold-one public-good dilemma: everyone benefits from provision, but each prefers another player to bear its cost. Unlike a dominant-defection dilemma, the best response depends on whether another player volunteers. Since $0<c<B$ throughout our grid, the pure-strategy Nash equilibria have exactly one volunteer. The cooperation rate here measures willingness to supply the benefit; group performance depends on securing provision while avoiding redundant costs.

\subsection{Assurance}
\label{app:payoffs-assurance}
Three players choose to join a project ($C$) or take a safe outside option ($D$). A contributor receives $h$ if at least $k\in\{2,3\}$ players, including itself, contribute, and zero otherwise. Opting out always gives 3, including when the others complete the project. Thus
\[
u_i=\begin{cases}h\mathbf1\{m_{-i}+1\ge k\},&a_i=1,\\3,&a_i=0,\end{cases}
\qquad h=3(1+\lambda),\quad \lambda\in\{0.05,0.10,0.20,0.40\}.
\]
Here $h$ is the contributor's total payoff when the project succeeds.

\begin{center}
\begin{minipage}{\linewidth}
\centering\small
\textbf{General payoff matrix: focal payoff $u_i$, $k\in\{2,3\}$.}\par\smallskip
\begin{tabular}{@{}lccc@{}}
\toprule
Own action & $m_{-i}=0$ & $m_{-i}=1$ & $m_{-i}=2$ \\
\midrule
$C$: join & $0$ & $h\mathbf1\{k=2\}$ & $h$ \\
$D$: safe option & $3$ & $3$ & $3$ \\
\bottomrule
\end{tabular}
\par\medskip
\textbf{Example: $k=2$, $\lambda=0.2$, hence $h=3.6$.}\par\smallskip
\begin{tabular}{@{}lccc@{}}
\toprule
Own action & $m_{-i}=0$ & $m_{-i}=1$ & $m_{-i}=2$ \\
\midrule
$C$: join & $0$ & $3.6$ & $3.6$ \\
$D$: safe option & $3$ & $3$ & $3$ \\
\bottomrule
\end{tabular}
\end{minipage}
\end{center}
In this example, $(C,C,D)$ gives $(3.6,3.6,3)$, whereas $(C,D,D)$ gives $(0,3,3)$. At threshold $k=3$ with the same $\lambda$, the contributor's middle-column payoff becomes zero; the other entries are unchanged.

Full cooperation is a strict stage-game Nash equilibrium throughout this grid: against two contributors, joining pays $h>3$, more than the outside option. All opting out is also a strict equilibrium because one participant cannot meet either threshold. The cooperative outcome makes every player better off, and reaching it requires coordinating participation.

\subsection{Traveler's Dilemma (TD)}
\label{app:payoffs-td}
Two players choose integer claims from 2 to $K+1$, with $K\in\{8,16,32,64\}$ legal levels. Equal claims pay that claim to both players. For unequal claims, the lower claimant receives its claim plus bonus $B$, while the higher claimant receives the lower claim minus $B$, where $B=1+\tau(K-1)$ and $\tau\in\{0.025,0.05,0.10,0.20\}$. For a chosen claim $c$, the reported behavioral outcome is $(c-2)/(K-1)$; this normalized claim is distinct from payoff.

\begin{center}
\begin{minipage}{\linewidth}
\centering\small
\textbf{General payoff matrix: entries $(u_1,u_2)$.}\par
Rows are Player 1's claims; columns are Player 2's claims.\par\smallskip
\begin{tabular}{@{}lcccc@{}}
\toprule
Claim & $2$ & $3$ & $\cdots$ & $K+1$ \\
\midrule
$2$ & $(2,2)$ & $(2+B,2-B)$ & $\cdots$ & $(2+B,2-B)$ \\
$3$ & $(2-B,2+B)$ & $(3,3)$ & $\cdots$ & $(3+B,3-B)$ \\
$\vdots$ & $\vdots$ & $\vdots$ & $\ddots$ & $\vdots$ \\
$K+1$ & $(2-B,2+B)$ & $(3-B,3+B)$ & $\cdots$ & $(K+1,K+1)$ \\
\bottomrule
\end{tabular}
\par\medskip
For $K=8$ and $\tau=0.1$, $B=1.7$: equal claims of 9 pay $(9,9)$, while claims $(8,9)$ pay $(9.7,6.3)$. Undercutting raises the lower claimant's payoff by $0.7$, and the general matrix specifies all 64 joint actions.
\end{minipage}
\end{center}

\subsection{Trust}
\label{app:payoffs-trust}
Player 1 is an investor with endowment $E=10$; Players 2 and 3 are trustees. Keeping the endowment gives $(10,0,0)$. Investing creates $G=5m$ for each trustee, with multiplier $m\in\{2,3,4\}$. Each trustee's contingent action either returns $r=5(1+s)$ to the investor ($C$) or retains everything ($D$), with $s\in\{0.2,0.4,0.6,0.8\}$. Returning therefore leaves a trustee $G-r$, while retaining gives $G$, conditional on investment. The investor receives the sum returned by the two trustees.

\begin{center}
\begin{minipage}{\linewidth}
\centering\small
\textbf{General payoff matrix: entries $(u_1,u_2,u_3)$.}\par
Columns give the ordered trustee actions $(a_2,a_3)$.\par\smallskip
\setlength{\tabcolsep}{4pt}
\begin{tabular}{@{}lcccc@{}}
\toprule
Investor action & $(C,C)$ & $(C,D)$ & $(D,C)$ & $(D,D)$ \\
\midrule
$C$: invest & $(2r,G-r,G-r)$ & $(r,G-r,G)$ & $(r,G,G-r)$ & $(0,G,G)$ \\
$D$: keep & $(10,0,0)$ & $(10,0,0)$ & $(10,0,0)$ & $(10,0,0)$ \\
\bottomrule
\end{tabular}
\par\medskip
\textbf{Example: $m=3$, $s=0.4$, hence $G=15$ and $r=7$.}\par\smallskip
\begin{tabular}{@{}lcccc@{}}
\toprule
Investor action & $(C,C)$ & $(C,D)$ & $(D,C)$ & $(D,D)$ \\
\midrule
$C$: invest & $(14,8,8)$ & $(7,8,15)$ & $(7,15,8)$ & $(0,15,15)$ \\
$D$: keep & $(10,0,0)$ & $(10,0,0)$ & $(10,0,0)$ & $(10,0,0)$ \\
\bottomrule
\end{tabular}
\end{minipage}
\end{center}
The implementation elicits all three intentions simultaneously and executes trustee returns conditional on investment.

\FloatBarrier

\section{History Effects by Model Lineage}
\label{app:model-first}
Table~\ref{tab:model-first-history} expands the six-binary-game history comparison in Section~\ref{sec:f2} by lineage. Within each lineage, we average the five trajectory seeds and payoff configurations in each game, then weight games equally. The table reports HL40 minus HL2 at each training stage and the corresponding paired differences between stages, in pp. Negative $H_s$ means cooperation declines with longer history; negative $K_{s/t}$ means a larger decline, or smaller increase, at stage $s$ than at stage $t$ (Equation~\ref{eq:history}). InternLM2's SFT and Aligned declines are smaller than Base's.
\begin{table}[!htbp]
\centering\footnotesize
\setlength{\tabcolsep}{2.5pt}
\caption{History changes for each lineage, averaged over the six binary games. Each entry is mean $\pm$ SD in pp across five matched run IDs, after averaging payoff configurations within game and games equally. The SD summarizes variation among repeated runs within each lineage.}\label{tab:model-first-history}
\begin{tabular}{lrrrrrr}\toprule
Lineage & $H_B$ & $H_S$ & $H_A$ & $K_{S/B}$ & $K_{A/B}$ & $K_{A/S}$ \\
\midrule
\csmodel{ai2}{OLMo} & $+1.7\pm1.4$ & $-13.5\pm1.4$ & $-8.9\pm0.8$ & $-15.2\pm2.8$ & $-10.7\pm2.0$ & $+4.6\pm1.0$ \\
\csmodel{huggingface}{SmolLM2} & $+0.0\pm0.1$ & $-16.1\pm2.4$ & $-18.1\pm2.3$ & $-16.2\pm2.4$ & $-18.1\pm2.4$ & $-1.9\pm2.0$ \\
\csmodel{deepseek}{DeepSeekMath} & $+0.9\pm0.2$ & $-5.6\pm0.8$ & $-9.7\pm1.1$ & $-6.5\pm0.8$ & $-10.7\pm1.0$ & $-4.1\pm1.2$ \\
\csmodel{internlm}{InternLM2} & $-7.6\pm0.8$ & $-2.3\pm0.4$ & $-2.8\pm0.5$ & $+5.3\pm0.9$ & $+4.8\pm0.6$ & $-0.5\pm0.6$ \\
\csmodel{nvidia}{Nemotron} & $-6.0\pm1.7$ & $-10.9\pm1.5$ & $-19.7\pm2.7$ & $-4.9\pm2.9$ & $-13.7\pm3.8$ & $-8.8\pm3.0$ \\
\csmodel{tulu}{Tulu} & $+1.4\pm0.8$ & $-6.0\pm0.9$ & $-12.9\pm1.1$ & $-7.4\pm0.9$ & $-14.3\pm0.3$ & $-6.9\pm1.0$ \\
\csmodel{zephyr}{Zephyr} & $+1.3\pm0.6$ & $-2.4\pm0.2$ & $-4.4\pm0.8$ & $-3.7\pm0.6$ & $-5.7\pm1.0$ & $-2.0\pm0.7$ \\
\bottomrule\end{tabular}
\end{table}

\FloatBarrier

\section{Payoff Responses by Model Lineage}
\label{app:centered}
Table~\ref{tab:surfaces-full} adds paired stage differences to the four-point payoff summaries in the main text. Tables~\ref{tab:payoff-lineage-pgg}--\ref{tab:payoff-lineage-td} retain each lineage separately, with SD across five repeated runs. Within each run, we average the other payoff parameters and fit all four values of the game's main parameter. Stage differences pair runs by seed ID; realized random draws can differ across deployments.

For a game's four tested raw parameter values $x_1,\ldots,x_4$, ordered from $D_1$ to $D_4$, let $z_j=(x_j-x_1)/(x_4-x_1)$. This preserves the numerical spacing while setting $z_1=0$ and $z_4=1$, including when the raw values descend. Fit the cooperation rate as a fraction with an intercept, $\hat f(z)=\hat\eta+\hat\beta_z z$. The fitted drop in percentage points is
\begin{equation}
S=100[\hat f(0)-\hat f(1)]=-100\hat\beta_z.
\label{eq:S}
\end{equation}
A positive $S$ is a downward fitted trend across the tested range; individual adjacent observations may still rise.

\begin{figure}[!htbp]
\centering
\includegraphics[width=\linewidth]{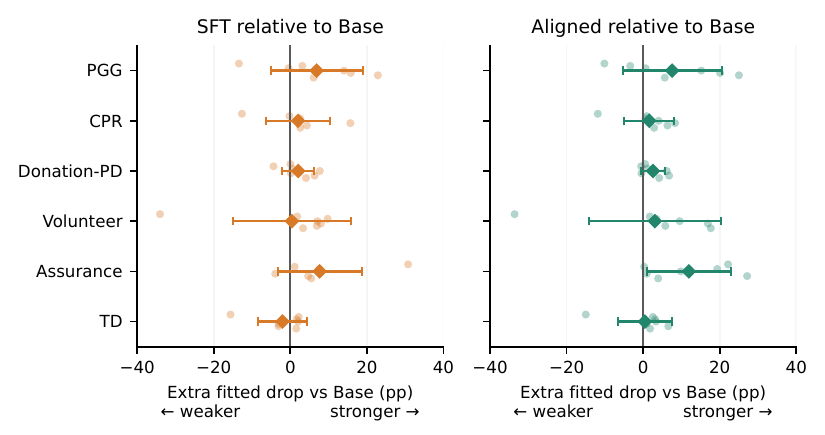}
\caption{Fitted payoff-drop differences relative to Base. Positive values mean a larger fitted decline after post-training across the tested payoff span. Each point is a lineage difference, averaging runs and the other payoff parameters; diamonds and bars show mean $\pm$ SD across seven lineages. Both panels use the same scale. TD measures normalized claims.}
\label{fig:centered}
\end{figure}

Models with high Base cooperation can also have high SFT cooperation. Pairing stages within lineage lets Figure~\ref{fig:centered} display their response differences. To compare variation between models with variation among payoff settings within a model, let $d_{\ell c}$ be the HL40 SFT-minus-Base difference in pp for lineage $\ell$ and configuration $c$ in one game, and let $\bar d_\ell$ be its configuration average. Then
\[
\operatorname{Var}_{\ell,c}(d_{\ell c})=
\operatorname{Var}_{\ell}(\bar d_\ell)+
E_\ell\operatorname{Var}_{c}(d_{\ell c}).
\]
With equal weights over observed estimates, differences between lineage averages account for 90.8\% of this variation in PGG and 89.8\% in CPR (Table~\ref{tab:variance}).
\begin{table}[t]
\centering
\small
\setlength{\tabcolsep}{3pt}
\caption{Where does variation in the HL40 SFT--Base difference come from? We separate differences between lineage averages from differences among payoff settings within a lineage. Entries use equal lineage/configuration weights and population variance, in $\mathrm{pp}^2$, calculated from the observed estimates.}\label{tab:variance}
\begin{tabular}{lrrr}
\toprule
Game & Total & Between lineage & Within lineage/config \\
\midrule
PGG & 1586.55 & 1439.81 & 146.74 \\
CPR & 829.71 & 745.27 & 84.44 \\
\bottomrule
\end{tabular}
\end{table}

\FloatBarrier
\begin{table}[!ht]
\centering
\setlength{\tabcolsep}{3pt}
\caption{\textbf{Full HL40 payoff surfaces and paired stage differences.} Each game's first row gives four values of its main payoff parameter ($D_1$--$D_4$), ordered toward higher cooperation costs, lower benefits, or greater rewards for noncooperation. Entries average seven lineages equally, after averaging seeds and secondary parameters within lineage. Stage rows give cooperation rates (\%; TD normalized claims multiplied by 100); difference rows give SFT minus Base and Aligned minus Base (pp). Negative differences mean lower cooperation than Base (lower claims in TD). When a negative difference becomes more negative from $D_1$ to $D_4$, the post-trained stage falls further below Base. The final column uses a straight-line fit to all four points (Equation~\ref{eq:S}). In a stage row, a positive drop means a fitted decline; in a difference row, it is the post-trained drop minus Base's, so a positive value means a larger fitted decline than Base's.}\label{tab:surfaces-full}
\small
\begin{tabular}{@{}lrrrrr@{}}\toprule
Game / stage & $D_1$ & $D_2$ & $D_3$ & $D_4$ & Fitted drop \\
\midrule
\textbf{PGG}: MPCR & 0.8 & 0.6 & 0.5 & 0.4 & -- \\
\cmidrule(l){1-6}
Base (B) & $59.8$ & $58.6$ & $57.1$ & $57.2$ & $+2.9$ \\
SFT (S) & $43.2$ & $38.6$ & $35.4$ & $33.7$ & $+9.8$ \\
Aligned (A) & $42.2$ & $38.7$ & $35.6$ & $31.4$ & $+10.5$ \\
SFT $-$ Base & $-16.6$ & $-20.0$ & $-21.6$ & $-23.5$ & $+6.9$ \\
Aligned $-$ Base & $-17.7$ & $-19.9$ & $-21.5$ & $-25.8$ & $+7.6$ \\
\midrule
\textbf{CPR}: $\theta$ & 0.2 & 0.4 & 0.6 & 0.8 & -- \\
\cmidrule(l){1-6}
Base (B) & $68.0$ & $63.7$ & $62.3$ & $65.6$ & $+2.6$ \\
SFT (S) & $31.4$ & $33.2$ & $30.5$ & $27.1$ & $+4.6$ \\
Aligned (A) & $23.6$ & $26.2$ & $24.6$ & $19.5$ & $+4.2$ \\
SFT $-$ Base & $-36.6$ & $-30.5$ & $-31.8$ & $-38.5$ & $+2.0$ \\
Aligned $-$ Base & $-44.4$ & $-37.5$ & $-37.7$ & $-46.1$ & $+1.6$ \\
\midrule
\textbf{Donation-PD}: $R$ & 2.4 & 2 & 1.4 & 1.1 & -- \\
\cmidrule(l){1-6}
Base (B) & $54.3$ & $52.9$ & $53.7$ & $54.3$ & $-0.2$ \\
SFT (S) & $15.9$ & $12.6$ & $13.6$ & $13.4$ & $+1.8$ \\
Aligned (A) & $11.7$ & $8.9$ & $9.3$ & $8.7$ & $+2.3$ \\
SFT $-$ Base & $-38.4$ & $-40.3$ & $-40.1$ & $-41.0$ & $+2.1$ \\
Aligned $-$ Base & $-42.7$ & $-44.1$ & $-44.4$ & $-45.6$ & $+2.5$ \\
\midrule
\textbf{Volunteer}: $\rho$ & 0.125 & 0.375 & 0.625 & 0.875 & -- \\
\cmidrule(l){1-6}
Base (B) & $80.6$ & $76.6$ & $75.1$ & $70.2$ & $+9.8$ \\
SFT (S) & $51.9$ & $49.3$ & $44.2$ & $42.3$ & $+10.2$ \\
Aligned (A) & $46.0$ & $42.6$ & $36.5$ & $33.8$ & $+12.8$ \\
SFT $-$ Base & $-28.6$ & $-27.2$ & $-30.9$ & $-27.8$ & $+0.4$ \\
Aligned $-$ Base & $-34.5$ & $-34.0$ & $-38.6$ & $-36.4$ & $+3.1$ \\
\midrule
\textbf{Assurance}: $\lambda$ & 0.4 & 0.2 & 0.1 & 0.05 & -- \\
\cmidrule(l){1-6}
Base (B) & $87.2$ & $85.7$ & $86.8$ & $87.3$ & $+0.0$ \\
SFT (S) & $80.8$ & $78.3$ & $71.5$ & $75.0$ & $+7.7$ \\
Aligned (A) & $85.2$ & $80.0$ & $70.4$ & $76.4$ & $+12.0$ \\
SFT $-$ Base & $-6.4$ & $-7.3$ & $-15.4$ & $-12.3$ & $+7.7$ \\
Aligned $-$ Base & $-2.0$ & $-5.7$ & $-16.5$ & $-10.9$ & $+11.9$ \\
\midrule
\textbf{TD}: $\tau$ & 0.025 & 0.05 & 0.1 & 0.2 & -- \\
\cmidrule(l){1-6}
Base (B) & $39.6$ & $39.8$ & $40.7$ & $37.9$ & $+1.7$ \\
SFT (S) & $67.2$ & $66.9$ & $67.0$ & $67.4$ & $-0.3$ \\
Aligned (A) & $62.6$ & $62.1$ & $61.4$ & $60.4$ & $+2.2$ \\
SFT $-$ Base & $+27.6$ & $+27.1$ & $+26.4$ & $+29.5$ & $-2.0$ \\
Aligned $-$ Base & $+23.0$ & $+22.4$ & $+20.8$ & $+22.4$ & $+0.4$ \\
\bottomrule\end{tabular}
\end{table}

\FloatBarrier
\begingroup
\makeatletter
\let\payoffSavedLTCaption\LT@makecaption
\long\def\LT@makecaption#1#2#3{\payoffSavedLTCaption#1{\normalsize#2}{\normalsize#3}}
\makeatother
\begingroup
\setlength{\LTcapwidth}{\linewidth}
\setlength{\tabcolsep}{4pt}
\small
\begin{longtable}{@{}lrrrrr@{}}
\caption{\textbf{PGG: HL40 payoff responses for each lineage.} Entries are mean $\pm$ SD across five runs, averaging secondary parameters within each run. Cooperation rates are percentages; stage differences and fitted declines are in pp. Stage differences match runs by seed ID. The final column fits all four parameter values within each run before taking the mean and SD.}\label{tab:payoff-lineage-pgg}\\
\toprule
Lineage / stage & $D_1$ & $D_2$ & $D_3$ & $D_4$ & Fitted drop \\
\midrule
\textbf{PGG}: MPCR & 0.8 & 0.6 & 0.5 & 0.4 & -- \\
\midrule
\endfirsthead
\multicolumn{6}{c}{\tablename\ \thetable{}: PGG (continued)}\\
\toprule
Lineage / stage & $D_1$ & $D_2$ & $D_3$ & $D_4$ & Fitted drop \\
\midrule
\textbf{PGG}: MPCR & 0.8 & 0.6 & 0.5 & 0.4 & -- \\
\midrule
\endhead
\midrule\multicolumn{6}{r}{Continued on next page}\\\endfoot
\bottomrule\endlastfoot
\multicolumn{6}{@{}l}{\textbf{OLMo-3-7B}}\\*
Base (B) & $20.3\pm23.7$ & $10.1\pm7.3$ & $3.3\pm2.6$ & $4.8\pm1.3$ & $+17.0\pm24.9$ \\*
SFT (S) & $8.5\pm2.1$ & $7.6\pm3.0$ & $6.2\pm1.5$ & $4.9\pm0.4$ & $+3.6\pm2.1$ \\*
Aligned (A) & $16.1\pm6.1$ & $13.4\pm3.2$ & $11.7\pm2.9$ & $9.0\pm2.5$ & $+6.8\pm4.5$ \\*
SFT $-$ Base & $-11.8\pm23.9$ & $-2.4\pm9.3$ & $+2.9\pm2.6$ & $+0.1\pm1.5$ & $-13.4\pm25.5$ \\*
Aligned $-$ Base & $-4.1\pm28.5$ & $+3.3\pm8.3$ & $+8.4\pm3.8$ & $+4.2\pm2.4$ & $-10.1\pm27.8$ \\
\midrule
\multicolumn{6}{@{}l}{\textbf{SmolLM2-1.7B}}\\*
Base (B) & $99.8\pm0.1$ & $99.8\pm0.1$ & $100.0\pm0.1$ & $99.9\pm0.1$ & $-0.1\pm0.2$ \\*
SFT (S) & $40.7\pm7.0$ & $34.6\pm3.0$ & $37.8\pm7.2$ & $37.3\pm5.9$ & $+3.1\pm1.5$ \\*
Aligned (A) & $34.1\pm4.8$ & $34.8\pm3.2$ & $39.4\pm6.0$ & $36.3\pm6.2$ & $-3.5\pm7.4$ \\*
SFT $-$ Base & $-59.1\pm6.9$ & $-65.1\pm2.9$ & $-62.2\pm7.1$ & $-62.6\pm5.9$ & $+3.1\pm1.5$ \\*
Aligned $-$ Base & $-65.8\pm4.8$ & $-64.9\pm3.1$ & $-60.6\pm6.0$ & $-63.6\pm6.1$ & $-3.4\pm7.4$ \\
\midrule
\multicolumn{6}{@{}l}{\textbf{DeepSeekMath-7B}}\\*
Base (B) & $98.6\pm0.2$ & $98.9\pm0.2$ & $98.9\pm0.4$ & $98.8\pm0.3$ & $-0.2\pm0.3$ \\*
SFT (S) & $35.6\pm6.6$ & $34.4\pm6.4$ & $37.5\pm9.0$ & $35.4\pm4.8$ & $-0.7\pm9.1$ \\*
Aligned (A) & $3.8\pm0.7$ & $3.8\pm0.7$ & $4.3\pm1.7$ & $3.1\pm0.6$ & $+0.4\pm1.2$ \\*
SFT $-$ Base & $-63.1\pm6.6$ & $-64.4\pm6.4$ & $-61.3\pm8.9$ & $-63.4\pm4.9$ & $-0.5\pm9.3$ \\*
Aligned $-$ Base & $-94.8\pm0.8$ & $-95.1\pm0.8$ & $-94.5\pm1.5$ & $-95.7\pm0.6$ & $+0.6\pm1.3$ \\
\midrule
\multicolumn{6}{@{}l}{\textbf{InternLM2-7B}}\\*
Base (B) & $36.9\pm2.1$ & $37.2\pm1.5$ & $35.3\pm2.1$ & $37.9\pm2.4$ & $-0.3\pm2.5$ \\*
SFT (S) & $46.4\pm2.9$ & $43.3\pm2.9$ & $33.4\pm1.3$ & $34.4\pm2.6$ & $+13.8\pm4.4$ \\*
Aligned (A) & $54.4\pm0.9$ & $48.7\pm1.9$ & $41.0\pm2.7$ & $40.7\pm2.8$ & $+14.9\pm2.9$ \\*
SFT $-$ Base & $+9.6\pm4.9$ & $+6.1\pm3.9$ & $-1.9\pm3.3$ & $-3.5\pm4.4$ & $+14.0\pm6.3$ \\*
Aligned $-$ Base & $+17.6\pm2.2$ & $+11.5\pm1.7$ & $+5.8\pm4.1$ & $+2.8\pm3.5$ & $+15.2\pm1.4$ \\
\midrule
\multicolumn{6}{@{}l}{\textbf{Nemotron-Cascade-8B}}\\*
Base (B) & $16.8\pm19.6$ & $17.7\pm16.4$ & $14.7\pm12.0$ & $12.8\pm2.6$ & $+4.0\pm16.5$ \\*
SFT (S) & $63.9\pm14.2$ & $56.3\pm10.2$ & $45.2\pm22.1$ & $46.1\pm9.0$ & $+19.8\pm20.0$ \\*
Aligned (A) & $50.7\pm9.7$ & $52.9\pm9.7$ & $32.4\pm20.1$ & $28.7\pm5.1$ & $+24.1\pm10.3$ \\*
SFT $-$ Base & $+47.2\pm26.9$ & $+38.7\pm19.3$ & $+30.6\pm30.7$ & $+33.3\pm8.6$ & $+15.8\pm29.9$ \\*
Aligned $-$ Base & $+33.9\pm24.0$ & $+35.2\pm23.8$ & $+17.8\pm29.8$ & $+15.9\pm4.9$ & $+20.1\pm20.8$ \\
\midrule
\multicolumn{6}{@{}l}{\textbf{T\"ulu-3-8B}}\\*
Base (B) & $98.5\pm0.1$ & $98.5\pm0.5$ & $98.5\pm0.4$ & $98.4\pm0.3$ & $+0.1\pm0.2$ \\*
SFT (S) & $50.6\pm2.6$ & $37.7\pm4.8$ & $36.1\pm2.8$ & $26.2\pm1.0$ & $+23.0\pm1.5$ \\*
Aligned (A) & $45.4\pm6.8$ & $26.9\pm2.4$ & $29.8\pm5.6$ & $18.0\pm1.7$ & $+25.1\pm8.6$ \\*
SFT $-$ Base & $-48.0\pm2.7$ & $-60.7\pm5.1$ & $-62.4\pm3.0$ & $-72.2\pm1.1$ & $+22.9\pm1.5$ \\*
Aligned $-$ Base & $-53.1\pm6.7$ & $-71.6\pm2.9$ & $-68.7\pm6.0$ & $-80.4\pm1.9$ & $+25.0\pm8.8$ \\
\midrule
\multicolumn{6}{@{}l}{\textbf{Zephyr-7B}}\\*
Base (B) & $47.9\pm1.5$ & $47.9\pm2.1$ & $48.8\pm1.9$ & $47.9\pm0.9$ & $-0.3\pm1.4$ \\*
SFT (S) & $56.8\pm1.5$ & $55.9\pm3.3$ & $51.8\pm1.2$ & $51.6\pm1.4$ & $+5.8\pm1.4$ \\*
Aligned (A) & $90.6\pm2.9$ & $90.5\pm2.3$ & $90.2\pm1.2$ & $84.1\pm1.9$ & $+5.3\pm2.9$ \\*
SFT $-$ Base & $+8.9\pm0.7$ & $+8.0\pm2.7$ & $+2.9\pm2.0$ & $+3.8\pm1.2$ & $+6.1\pm0.9$ \\*
Aligned $-$ Base & $+42.7\pm2.6$ & $+42.6\pm3.6$ & $+41.3\pm0.9$ & $+36.2\pm2.6$ & $+5.6\pm2.2$ \\
\end{longtable}
\endgroup
\begingroup
\setlength{\LTcapwidth}{\linewidth}
\setlength{\tabcolsep}{4pt}
\small
\begin{longtable}{@{}lrrrrr@{}}
\caption{\textbf{CPR: HL40 payoff responses for each lineage.} Entries are mean $\pm$ SD across five runs, averaging secondary parameters within each run. Cooperation rates are percentages; stage differences and fitted declines are in pp. Stage differences match runs by seed ID. The final column fits all four parameter values within each run before taking the mean and SD.}\label{tab:payoff-lineage-cpr}\\
\toprule
Lineage / stage & $D_1$ & $D_2$ & $D_3$ & $D_4$ & Fitted drop \\
\midrule
\textbf{CPR}: $\theta$ & 0.2 & 0.4 & 0.6 & 0.8 & -- \\
\midrule
\endfirsthead
\multicolumn{6}{c}{\tablename\ \thetable{}: CPR (continued)}\\
\toprule
Lineage / stage & $D_1$ & $D_2$ & $D_3$ & $D_4$ & Fitted drop \\
\midrule
\textbf{CPR}: $\theta$ & 0.2 & 0.4 & 0.6 & 0.8 & -- \\
\midrule
\endhead
\midrule\multicolumn{6}{r}{Continued on next page}\\\endfoot
\bottomrule\endlastfoot
\multicolumn{6}{@{}l}{\textbf{OLMo-3-7B}}\\*
Base (B) & $56.5\pm8.7$ & $45.9\pm7.1$ & $45.5\pm8.4$ & $42.6\pm9.4$ & $+12.7\pm8.8$ \\*
SFT (S) & $12.3\pm2.5$ & $12.6\pm3.7$ & $13.0\pm6.3$ & $12.1\pm2.8$ & $+0.0\pm2.8$ \\*
Aligned (A) & $9.8\pm1.6$ & $7.9\pm0.7$ & $8.2\pm0.3$ & $8.8\pm1.5$ & $+0.8\pm1.9$ \\*
SFT $-$ Base & $-44.2\pm6.6$ & $-33.3\pm4.4$ & $-32.5\pm10.6$ & $-30.4\pm10.8$ & $-12.7\pm11.6$ \\*
Aligned $-$ Base & $-46.7\pm8.2$ & $-38.0\pm6.5$ & $-37.3\pm8.3$ & $-33.8\pm10.7$ & $-11.9\pm8.5$ \\
\midrule
\multicolumn{6}{@{}l}{\textbf{SmolLM2-1.7B}}\\*
Base (B) & $99.6\pm0.1$ & $99.6\pm0.2$ & $99.6\pm0.2$ & $99.7\pm0.1$ & $-0.1\pm0.1$ \\*
SFT (S) & $8.5\pm1.5$ & $19.0\pm2.7$ & $17.1\pm3.2$ & $9.5\pm1.4$ & $-0.3\pm0.9$ \\*
Aligned (A) & $8.4\pm0.9$ & $12.9\pm2.7$ & $12.6\pm1.4$ & $7.6\pm0.8$ & $+0.8\pm1.5$ \\*
SFT $-$ Base & $-91.0\pm1.5$ & $-80.6\pm2.6$ & $-82.4\pm3.2$ & $-90.2\pm1.3$ & $-0.2\pm1.0$ \\*
Aligned $-$ Base & $-91.2\pm0.8$ & $-86.6\pm2.7$ & $-87.0\pm1.3$ & $-92.1\pm0.7$ & $+0.9\pm1.5$ \\
\midrule
\multicolumn{6}{@{}l}{\textbf{DeepSeekMath-7B}}\\*
Base (B) & $97.8\pm0.4$ & $83.4\pm3.2$ & $83.7\pm1.9$ & $97.8\pm0.4$ & $-0.0\pm0.9$ \\*
SFT (S) & $54.1\pm3.7$ & $66.3\pm3.0$ & $63.3\pm3.1$ & $52.3\pm2.8$ & $+2.5\pm5.5$ \\*
Aligned (A) & $8.8\pm1.1$ & $36.5\pm2.8$ & $34.4\pm4.2$ & $8.9\pm0.7$ & $+0.5\pm0.8$ \\*
SFT $-$ Base & $-43.8\pm3.9$ & $-17.1\pm2.0$ & $-20.4\pm3.5$ & $-45.5\pm3.1$ & $+2.6\pm5.7$ \\*
Aligned $-$ Base & $-89.0\pm0.8$ & $-46.9\pm4.0$ & $-49.3\pm4.2$ & $-88.9\pm0.8$ & $+0.6\pm1.2$ \\
\midrule
\multicolumn{6}{@{}l}{\textbf{InternLM2-7B}}\\*
Base (B) & $45.2\pm1.7$ & $43.6\pm1.6$ & $42.3\pm1.3$ & $45.5\pm1.0$ & $+0.1\pm1.0$ \\*
SFT (S) & $27.9\pm2.5$ & $25.0\pm1.0$ & $24.7\pm1.0$ & $25.8\pm1.0$ & $+2.0\pm2.2$ \\*
Aligned (A) & $40.3\pm2.0$ & $38.8\pm1.5$ & $36.5\pm1.7$ & $36.5\pm2.0$ & $+4.1\pm3.7$ \\*
SFT $-$ Base & $-17.3\pm2.9$ & $-18.5\pm1.7$ & $-17.7\pm1.5$ & $-19.7\pm1.9$ & $+1.9\pm2.0$ \\*
Aligned $-$ Base & $-4.9\pm1.7$ & $-4.8\pm2.1$ & $-5.9\pm1.9$ & $-9.1\pm2.8$ & $+4.0\pm3.0$ \\
\midrule
\multicolumn{6}{@{}l}{\textbf{Nemotron-Cascade-8B}}\\*
Base (B) & $22.0\pm10.4$ & $35.3\pm8.8$ & $25.8\pm4.9$ & $19.7\pm2.3$ & $+4.8\pm8.2$ \\*
SFT (S) & $46.4\pm7.0$ & $39.4\pm7.4$ & $30.4\pm9.1$ & $26.6\pm5.6$ & $+20.5\pm8.1$ \\*
Aligned (A) & $32.1\pm15.4$ & $22.3\pm2.3$ & $21.4\pm3.0$ & $17.8\pm3.5$ & $+13.2\pm15.4$ \\*
SFT $-$ Base & $+24.4\pm15.9$ & $+4.2\pm11.2$ & $+4.6\pm11.6$ & $+6.8\pm5.5$ & $+15.7\pm15.2$ \\*
Aligned $-$ Base & $+10.2\pm18.4$ & $-13.0\pm10.2$ & $-4.4\pm2.9$ & $-1.9\pm4.3$ & $+8.3\pm17.4$ \\
\midrule
\multicolumn{6}{@{}l}{\textbf{T\"ulu-3-8B}}\\*
Base (B) & $91.6\pm3.0$ & $75.3\pm2.9$ & $76.7\pm2.1$ & $90.5\pm4.7$ & $+0.5\pm2.1$ \\*
SFT (S) & $34.5\pm1.3$ & $34.3\pm2.0$ & $31.9\pm1.3$ & $29.9\pm1.7$ & $+4.9\pm2.0$ \\*
Aligned (A) & $33.3\pm1.2$ & $32.6\pm0.9$ & $28.4\pm1.0$ & $27.1\pm0.4$ & $+6.9\pm1.5$ \\*
SFT $-$ Base & $-57.1\pm2.4$ & $-41.0\pm2.8$ & $-44.9\pm1.9$ & $-60.6\pm5.7$ & $+4.3\pm4.0$ \\*
Aligned $-$ Base & $-58.2\pm2.9$ & $-42.7\pm2.7$ & $-48.4\pm2.1$ & $-63.4\pm4.8$ & $+6.3\pm2.6$ \\
\midrule
\multicolumn{6}{@{}l}{\textbf{Zephyr-7B}}\\*
Base (B) & $63.3\pm3.0$ & $62.8\pm1.8$ & $62.5\pm1.3$ & $63.2\pm2.5$ & $+0.1\pm3.6$ \\*
SFT (S) & $36.0\pm1.0$ & $35.5\pm2.3$ & $33.4\pm2.2$ & $33.7\pm2.5$ & $+2.7\pm1.9$ \\*
Aligned (A) & $32.6\pm1.2$ & $32.3\pm1.5$ & $30.7\pm1.2$ & $29.8\pm1.6$ & $+3.0\pm1.2$ \\*
SFT $-$ Base & $-27.2\pm2.9$ & $-27.3\pm2.1$ & $-29.0\pm1.9$ & $-29.6\pm2.9$ & $+2.6\pm2.5$ \\*
Aligned $-$ Base & $-30.7\pm2.9$ & $-30.6\pm3.1$ & $-31.8\pm1.7$ & $-33.5\pm2.5$ & $+2.8\pm4.6$ \\
\end{longtable}
\endgroup
\begingroup
\setlength{\LTcapwidth}{\linewidth}
\setlength{\tabcolsep}{4pt}
\small
\begin{longtable}{@{}lrrrrr@{}}
\caption{\textbf{Donation-PD: HL40 payoff responses for each lineage.} Entries are mean $\pm$ SD across five runs, averaging secondary parameters within each run. Cooperation rates are percentages; stage differences and fitted declines are in pp. Stage differences match runs by seed ID. The final column fits all four parameter values within each run before taking the mean and SD.}\label{tab:payoff-lineage-donationpd}\\
\toprule
Lineage / stage & $D_1$ & $D_2$ & $D_3$ & $D_4$ & Fitted drop \\
\midrule
\textbf{Donation-PD}: $R$ & 2.4 & 2 & 1.4 & 1.1 & -- \\
\midrule
\endfirsthead
\multicolumn{6}{c}{\tablename\ \thetable{}: Donation-PD (continued)}\\
\toprule
Lineage / stage & $D_1$ & $D_2$ & $D_3$ & $D_4$ & Fitted drop \\
\midrule
\textbf{Donation-PD}: $R$ & 2.4 & 2 & 1.4 & 1.1 & -- \\
\midrule
\endhead
\midrule\multicolumn{6}{r}{Continued on next page}\\\endfoot
\bottomrule\endlastfoot
\multicolumn{6}{@{}l}{\textbf{OLMo-3-7B}}\\*
Base (B) & $2.2\pm0.3$ & $1.7\pm0.5$ & $1.4\pm0.3$ & $1.4\pm0.3$ & $+0.8\pm0.4$ \\*
SFT (S) & $1.0\pm0.2$ & $0.4\pm0.2$ & $0.2\pm0.1$ & $0.1\pm0.0$ & $+0.8\pm0.2$ \\*
Aligned (A) & $1.7\pm0.1$ & $1.2\pm0.2$ & $0.6\pm0.1$ & $0.4\pm0.1$ & $+1.3\pm0.1$ \\*
SFT $-$ Base & $-1.3\pm0.4$ & $-1.3\pm0.6$ & $-1.2\pm0.3$ & $-1.3\pm0.3$ & $+0.0\pm0.5$ \\*
Aligned $-$ Base & $-0.5\pm0.2$ & $-0.5\pm0.5$ & $-0.8\pm0.4$ & $-1.0\pm0.4$ & $+0.5\pm0.4$ \\
\midrule
\multicolumn{6}{@{}l}{\textbf{SmolLM2-1.7B}}\\*
Base (B) & $99.7\pm0.1$ & $99.8\pm0.1$ & $99.7\pm0.1$ & $99.7\pm0.3$ & $+0.1\pm0.2$ \\*
SFT (S) & $7.8\pm3.4$ & $9.7\pm2.4$ & $11.7\pm4.7$ & $12.1\pm4.2$ & $-4.3\pm4.9$ \\*
Aligned (A) & $3.4\pm0.8$ & $5.6\pm1.7$ & $5.6\pm1.8$ & $3.8\pm1.0$ & $-0.5\pm2.0$ \\*
SFT $-$ Base & $-91.9\pm3.4$ & $-90.1\pm2.4$ & $-88.0\pm4.7$ & $-87.6\pm4.2$ & $-4.4\pm4.8$ \\*
Aligned $-$ Base & $-96.3\pm0.8$ & $-94.2\pm1.6$ & $-94.1\pm1.8$ & $-95.8\pm1.2$ & $-0.5\pm2.1$ \\
\midrule
\multicolumn{6}{@{}l}{\textbf{DeepSeekMath-7B}}\\*
Base (B) & $95.9\pm0.2$ & $96.1\pm0.6$ & $95.9\pm0.5$ & $96.0\pm0.7$ & $-0.0\pm0.8$ \\*
SFT (S) & $12.0\pm2.2$ & $5.8\pm0.9$ & $9.0\pm1.6$ & $10.2\pm1.7$ & $+0.5\pm1.1$ \\*
Aligned (A) & $5.8\pm0.7$ & $1.8\pm0.7$ & $3.5\pm1.0$ & $3.7\pm0.7$ & $+1.1\pm0.6$ \\*
SFT $-$ Base & $-83.9\pm2.2$ & $-90.3\pm1.1$ & $-86.9\pm1.9$ & $-85.8\pm1.6$ & $+0.5\pm1.4$ \\*
Aligned $-$ Base & $-90.2\pm0.5$ & $-94.2\pm1.0$ & $-92.4\pm1.1$ & $-92.3\pm1.2$ & $+1.1\pm1.1$ \\
\midrule
\multicolumn{6}{@{}l}{\textbf{InternLM2-7B}}\\*
Base (B) & $43.0\pm1.8$ & $41.7\pm1.2$ & $41.0\pm0.6$ & $42.7\pm1.5$ & $+0.6\pm1.8$ \\*
SFT (S) & $37.7\pm1.4$ & $28.7\pm1.1$ & $29.9\pm1.8$ & $27.3\pm1.1$ & $+8.3\pm0.6$ \\*
Aligned (A) & $41.1\pm2.5$ & $31.8\pm2.1$ & $32.7\pm1.6$ & $32.9\pm1.8$ & $+6.8\pm2.6$ \\*
SFT $-$ Base & $-5.3\pm2.1$ & $-13.0\pm1.3$ & $-11.1\pm1.3$ & $-15.4\pm2.6$ & $+7.7\pm2.1$ \\*
Aligned $-$ Base & $-1.9\pm3.0$ & $-9.9\pm3.3$ & $-8.3\pm1.4$ & $-9.9\pm2.5$ & $+6.2\pm3.1$ \\
\midrule
\multicolumn{6}{@{}l}{\textbf{Nemotron-Cascade-8B}}\\*
Base (B) & $1.0\pm0.3$ & $0.4\pm0.1$ & $0.7\pm0.3$ & $0.7\pm0.2$ & $+0.2\pm0.1$ \\*
SFT (S) & $5.5\pm1.7$ & $3.4\pm0.4$ & $3.8\pm0.5$ & $5.1\pm1.8$ & $+0.3\pm0.8$ \\*
Aligned (A) & $3.3\pm1.5$ & $1.6\pm0.3$ & $3.3\pm0.9$ & $2.9\pm0.5$ & $-0.3\pm1.3$ \\*
SFT $-$ Base & $+4.5\pm1.5$ & $+3.0\pm0.4$ & $+3.1\pm0.7$ & $+4.4\pm1.6$ & $+0.1\pm0.7$ \\*
Aligned $-$ Base & $+2.3\pm1.4$ & $+1.2\pm0.4$ & $+2.6\pm0.9$ & $+2.2\pm0.4$ & $-0.4\pm1.4$ \\
\midrule
\multicolumn{6}{@{}l}{\textbf{T\"ulu-3-8B}}\\*
Base (B) & $91.4\pm4.6$ & $81.1\pm8.6$ & $90.4\pm6.4$ & $92.6\pm4.2$ & $-4.2\pm4.5$ \\*
SFT (S) & $15.8\pm1.0$ & $14.2\pm0.6$ & $13.7\pm1.4$ & $13.5\pm0.7$ & $+2.2\pm0.7$ \\*
Aligned (A) & $7.3\pm1.5$ & $6.5\pm1.3$ & $4.8\pm0.9$ & $4.9\pm1.5$ & $+2.6\pm1.2$ \\*
SFT $-$ Base & $-75.5\pm4.7$ & $-66.9\pm8.8$ & $-76.7\pm6.7$ & $-79.1\pm4.3$ & $+6.3\pm4.4$ \\*
Aligned $-$ Base & $-84.1\pm3.4$ & $-74.5\pm8.4$ & $-85.6\pm6.0$ & $-87.7\pm4.1$ & $+6.8\pm4.7$ \\
\midrule
\multicolumn{6}{@{}l}{\textbf{Zephyr-7B}}\\*
Base (B) & $47.1\pm1.5$ & $50.0\pm2.3$ & $46.9\pm1.4$ & $47.2\pm2.1$ & $+1.0\pm2.5$ \\*
SFT (S) & $31.6\pm1.8$ & $26.1\pm1.0$ & $26.7\pm2.1$ & $25.3\pm1.5$ & $+5.0\pm1.3$ \\*
Aligned (A) & $19.0\pm3.4$ & $13.7\pm1.5$ & $14.9\pm0.9$ & $12.3\pm2.1$ & $+5.1\pm3.2$ \\*
SFT $-$ Base & $-15.5\pm1.6$ & $-23.8\pm2.4$ & $-20.2\pm3.2$ & $-21.9\pm3.1$ & $+4.1\pm3.2$ \\*
Aligned $-$ Base & $-28.1\pm4.6$ & $-36.2\pm2.4$ & $-32.0\pm1.2$ & $-35.0\pm2.0$ & $+4.2\pm3.8$ \\
\end{longtable}
\endgroup
\begingroup
\setlength{\LTcapwidth}{\linewidth}
\setlength{\tabcolsep}{4pt}
\small
\begin{longtable}{@{}lrrrrr@{}}
\caption{\textbf{Volunteer: HL40 payoff responses for each lineage.} Entries are mean $\pm$ SD across five runs, averaging secondary parameters within each run. Cooperation rates are percentages; stage differences and fitted declines are in pp. Stage differences match runs by seed ID. The final column fits all four parameter values within each run before taking the mean and SD.}\label{tab:payoff-lineage-volunteer}\\
\toprule
Lineage / stage & $D_1$ & $D_2$ & $D_3$ & $D_4$ & Fitted drop \\
\midrule
\textbf{Volunteer}: $\rho$ & 0.125 & 0.375 & 0.625 & 0.875 & -- \\
\midrule
\endfirsthead
\multicolumn{6}{c}{\tablename\ \thetable{}: Volunteer (continued)}\\
\toprule
Lineage / stage & $D_1$ & $D_2$ & $D_3$ & $D_4$ & Fitted drop \\
\midrule
\textbf{Volunteer}: $\rho$ & 0.125 & 0.375 & 0.625 & 0.875 & -- \\
\midrule
\endhead
\midrule\multicolumn{6}{r}{Continued on next page}\\\endfoot
\bottomrule\endlastfoot
\multicolumn{6}{@{}l}{\textbf{OLMo-3-7B}}\\*
Base (B) & $83.4\pm3.6$ & $59.5\pm15.2$ & $56.2\pm13.8$ & $30.3\pm3.7$ & $+48.8\pm7.4$ \\*
SFT (S) & $41.3\pm2.2$ & $35.8\pm0.7$ & $29.4\pm0.8$ & $27.0\pm1.6$ & $+14.8\pm2.0$ \\*
Aligned (A) & $45.2\pm2.4$ & $39.3\pm2.5$ & $34.6\pm3.6$ & $29.8\pm2.2$ & $+15.2\pm2.4$ \\*
SFT $-$ Base & $-42.1\pm4.3$ & $-23.7\pm15.3$ & $-26.8\pm14.1$ & $-3.2\pm4.2$ & $-34.0\pm5.7$ \\*
Aligned $-$ Base & $-38.3\pm2.9$ & $-20.2\pm14.4$ & $-21.6\pm10.9$ & $-0.5\pm4.0$ & $-33.6\pm5.8$ \\
\midrule
\multicolumn{6}{@{}l}{\textbf{SmolLM2-1.7B}}\\*
Base (B) & $99.6\pm0.2$ & $99.6\pm0.1$ & $99.6\pm0.1$ & $99.5\pm0.1$ & $+0.0\pm0.2$ \\*
SFT (S) & $38.8\pm6.4$ & $37.9\pm4.2$ & $36.2\pm6.8$ & $37.3\pm3.3$ & $+1.8\pm4.5$ \\*
Aligned (A) & $32.7\pm4.7$ & $28.5\pm2.9$ & $29.3\pm8.0$ & $30.5\pm6.8$ & $+1.8\pm10.7$ \\*
SFT $-$ Base & $-60.8\pm6.3$ & $-61.7\pm4.3$ & $-63.5\pm6.8$ & $-62.2\pm3.2$ & $+1.8\pm4.5$ \\*
Aligned $-$ Base & $-66.9\pm4.6$ & $-71.1\pm2.9$ & $-70.3\pm8.0$ & $-69.1\pm6.8$ & $+1.7\pm10.6$ \\
\midrule
\multicolumn{6}{@{}l}{\textbf{DeepSeekMath-7B}}\\*
Base (B) & $97.2\pm0.4$ & $97.3\pm0.2$ & $96.9\pm0.4$ & $96.2\pm0.2$ & $+1.0\pm0.4$ \\*
SFT (S) & $45.9\pm2.5$ & $43.8\pm3.4$ & $38.4\pm2.1$ & $35.8\pm1.6$ & $+10.7\pm1.7$ \\*
Aligned (A) & $16.5\pm1.0$ & $14.9\pm0.9$ & $13.7\pm0.4$ & $12.0\pm1.0$ & $+4.4\pm1.1$ \\*
SFT $-$ Base & $-51.3\pm2.5$ & $-53.5\pm3.4$ & $-58.5\pm1.9$ & $-60.4\pm1.8$ & $+9.7\pm1.8$ \\*
Aligned $-$ Base & $-80.7\pm0.9$ & $-82.4\pm1.0$ & $-83.1\pm0.8$ & $-84.2\pm1.0$ & $+3.4\pm0.7$ \\
\midrule
\multicolumn{6}{@{}l}{\textbf{InternLM2-7B}}\\*
Base (B) & $35.5\pm0.8$ & $36.0\pm1.3$ & $36.0\pm2.2$ & $36.1\pm1.4$ & $-0.5\pm1.7$ \\*
SFT (S) & $45.1\pm0.9$ & $43.5\pm0.9$ & $41.1\pm1.4$ & $38.6\pm1.6$ & $+6.6\pm2.3$ \\*
Aligned (A) & $53.8\pm2.2$ & $50.0\pm1.2$ & $45.7\pm1.0$ & $45.2\pm1.6$ & $+9.0\pm2.3$ \\*
SFT $-$ Base & $+9.6\pm0.9$ & $+7.5\pm1.8$ & $+5.2\pm3.4$ & $+2.5\pm1.9$ & $+7.1\pm2.4$ \\*
Aligned $-$ Base & $+18.3\pm2.7$ & $+13.9\pm1.5$ & $+9.7\pm1.8$ & $+9.1\pm2.9$ & $+9.5\pm3.7$ \\
\midrule
\multicolumn{6}{@{}l}{\textbf{Nemotron-Cascade-8B}}\\*
Base (B) & $98.3\pm1.7$ & $94.5\pm4.1$ & $89.1\pm8.0$ & $82.3\pm8.6$ & $+16.0\pm8.5$ \\*
SFT (S) & $86.5\pm3.6$ & $81.4\pm4.8$ & $68.1\pm2.3$ & $64.3\pm7.2$ & $+24.0\pm7.4$ \\*
Aligned (A) & $72.2\pm6.6$ & $69.2\pm5.7$ & $48.0\pm2.9$ & $42.8\pm1.7$ & $+32.9\pm6.9$ \\*
SFT $-$ Base & $-11.7\pm3.5$ & $-13.1\pm7.7$ & $-21.0\pm8.3$ & $-18.0\pm7.3$ & $+8.0\pm7.8$ \\*
Aligned $-$ Base & $-26.0\pm7.9$ & $-25.3\pm6.3$ & $-41.0\pm7.4$ & $-39.5\pm7.3$ & $+16.9\pm12.9$ \\
\midrule
\multicolumn{6}{@{}l}{\textbf{T\"ulu-3-8B}}\\*
Base (B) & $98.6\pm0.4$ & $98.0\pm0.5$ & $97.5\pm0.1$ & $97.1\pm0.5$ & $+1.5\pm0.8$ \\*
SFT (S) & $45.9\pm1.7$ & $44.1\pm1.2$ & $40.8\pm1.5$ & $37.6\pm0.8$ & $+8.4\pm1.7$ \\*
Aligned (A) & $37.9\pm1.5$ & $35.0\pm1.5$ & $33.6\pm2.3$ & $30.3\pm1.1$ & $+7.3\pm1.5$ \\*
SFT $-$ Base & $-52.7\pm1.9$ & $-54.0\pm1.3$ & $-56.8\pm1.5$ & $-59.5\pm0.6$ & $+6.9\pm1.7$ \\*
Aligned $-$ Base & $-60.7\pm1.6$ & $-63.0\pm1.9$ & $-63.9\pm2.4$ & $-66.8\pm1.5$ & $+5.8\pm1.9$ \\
\midrule
\multicolumn{6}{@{}l}{\textbf{Zephyr-7B}}\\*
Base (B) & $51.4\pm1.8$ & $51.0\pm0.9$ & $50.5\pm1.7$ & $49.8\pm1.6$ & $+1.5\pm1.2$ \\*
SFT (S) & $60.2\pm0.9$ & $58.8\pm2.1$ & $55.7\pm1.4$ & $55.9\pm1.5$ & $+4.9\pm2.1$ \\*
Aligned (A) & $63.9\pm5.0$ & $60.9\pm3.1$ & $50.3\pm1.9$ & $46.1\pm2.3$ & $+19.2\pm6.0$ \\*
SFT $-$ Base & $+8.9\pm2.0$ & $+7.8\pm2.2$ & $+5.1\pm2.6$ & $+6.0\pm2.3$ & $+3.3\pm2.3$ \\*
Aligned $-$ Base & $+12.6\pm5.8$ & $+9.9\pm3.4$ & $-0.2\pm2.1$ & $-3.7\pm1.4$ & $+17.7\pm5.8$ \\
\end{longtable}
\endgroup
\begingroup
\setlength{\LTcapwidth}{\linewidth}
\setlength{\tabcolsep}{4pt}
\small
\begin{longtable}{@{}lrrrrr@{}}
\caption{\textbf{Assurance: HL40 payoff responses for each lineage.} Entries are mean $\pm$ SD across five runs, averaging secondary parameters within each run. Cooperation rates are percentages; stage differences and fitted declines are in pp. Stage differences match runs by seed ID. The final column fits all four parameter values within each run before taking the mean and SD.}\label{tab:payoff-lineage-assurance}\\
\toprule
Lineage / stage & $D_1$ & $D_2$ & $D_3$ & $D_4$ & Fitted drop \\
\midrule
\textbf{Assurance}: $\lambda$ & 0.4 & 0.2 & 0.1 & 0.05 & -- \\
\midrule
\endfirsthead
\multicolumn{6}{c}{\tablename\ \thetable{}: Assurance (continued)}\\
\toprule
Lineage / stage & $D_1$ & $D_2$ & $D_3$ & $D_4$ & Fitted drop \\
\midrule
\textbf{Assurance}: $\lambda$ & 0.4 & 0.2 & 0.1 & 0.05 & -- \\
\midrule
\endhead
\midrule\multicolumn{6}{r}{Continued on next page}\\\endfoot
\bottomrule\endlastfoot
\multicolumn{6}{@{}l}{\textbf{OLMo-3-7B}}\\*
Base (B) & $98.3\pm0.6$ & $89.8\pm19.5$ & $98.0\pm0.2$ & $98.0\pm0.5$ & $-0.2\pm1.0$ \\*
SFT (S) & $89.9\pm18.2$ & $90.3\pm13.0$ & $45.6\pm16.5$ & $72.4\pm20.3$ & $+30.6\pm16.1$ \\*
Aligned (A) & $81.8\pm19.6$ & $71.3\pm22.7$ & $57.0\pm4.4$ & $63.8\pm18.8$ & $+22.0\pm30.0$ \\*
SFT $-$ Base & $-8.5\pm18.2$ & $+0.5\pm6.9$ & $-52.5\pm16.5$ & $-25.6\pm20.4$ & $+30.8\pm16.4$ \\*
Aligned $-$ Base & $-16.6\pm20.1$ & $-18.5\pm37.9$ & $-41.0\pm4.5$ & $-34.2\pm18.6$ & $+22.2\pm30.7$ \\
\midrule
\multicolumn{6}{@{}l}{\textbf{SmolLM2-1.7B}}\\*
Base (B) & $99.1\pm0.3$ & $99.1\pm0.4$ & $99.3\pm0.4$ & $99.5\pm0.4$ & $-0.3\pm0.2$ \\*
SFT (S) & $96.8\pm3.6$ & $93.8\pm6.5$ & $95.2\pm5.0$ & $96.2\pm3.6$ & $+0.8\pm3.2$ \\*
Aligned (A) & $97.6\pm4.0$ & $95.8\pm4.6$ & $96.7\pm3.7$ & $98.0\pm3.2$ & $-0.0\pm0.9$ \\*
SFT $-$ Base & $-2.4\pm3.5$ & $-5.3\pm6.1$ & $-4.1\pm4.7$ & $-3.2\pm3.2$ & $+1.1\pm3.3$ \\*
Aligned $-$ Base & $-1.6\pm3.8$ & $-3.3\pm4.1$ & $-2.6\pm3.4$ & $-1.4\pm2.9$ & $+0.3\pm1.0$ \\
\midrule
\multicolumn{6}{@{}l}{\textbf{DeepSeekMath-7B}}\\*
Base (B) & $99.2\pm0.3$ & $99.2\pm0.2$ & $99.0\pm0.3$ & $99.2\pm0.2$ & $+0.1\pm0.3$ \\*
SFT (S) & $87.3\pm2.3$ & $83.1\pm3.2$ & $77.0\pm3.6$ & $82.1\pm3.1$ & $+7.7\pm3.1$ \\*
Aligned (A) & $87.7\pm8.2$ & $78.0\pm9.5$ & $61.0\pm11.3$ & $74.7\pm18.1$ & $+19.4\pm5.1$ \\*
SFT $-$ Base & $-11.9\pm2.3$ & $-16.1\pm3.3$ & $-22.1\pm3.6$ & $-17.1\pm3.2$ & $+7.6\pm3.2$ \\*
Aligned $-$ Base & $-11.6\pm8.2$ & $-21.1\pm9.6$ & $-38.0\pm11.4$ & $-24.5\pm18.0$ & $+19.3\pm4.8$ \\
\midrule
\multicolumn{6}{@{}l}{\textbf{InternLM2-7B}}\\*
Base (B) & $40.6\pm2.1$ & $41.6\pm1.6$ & $40.2\pm1.9$ & $42.0\pm1.5$ & $-0.7\pm3.4$ \\*
SFT (S) & $71.1\pm2.8$ & $69.1\pm2.2$ & $64.3\pm2.9$ & $64.3\pm1.1$ & $+7.3\pm2.5$ \\*
Aligned (A) & $84.5\pm2.4$ & $82.2\pm1.8$ & $78.0\pm4.8$ & $74.8\pm3.3$ & $+9.1\pm4.4$ \\*
SFT $-$ Base & $+30.5\pm2.9$ & $+27.5\pm3.3$ & $+24.0\pm3.0$ & $+22.3\pm2.2$ & $+8.1\pm4.0$ \\*
Aligned $-$ Base & $+43.9\pm4.4$ & $+40.7\pm0.9$ & $+37.8\pm6.2$ & $+32.8\pm3.8$ & $+9.8\pm7.5$ \\
\midrule
\multicolumn{6}{@{}l}{\textbf{Nemotron-Cascade-8B}}\\*
Base (B) & $99.8\pm0.1$ & $99.7\pm0.2$ & $99.8\pm0.0$ & $99.8\pm0.1$ & $-0.1\pm0.2$ \\*
SFT (S) & $59.3\pm18.3$ & $52.1\pm1.4$ & $67.7\pm21.7$ & $59.3\pm35.3$ & $-4.0\pm21.9$ \\*
Aligned (A) & $70.8\pm25.3$ & $71.8\pm24.4$ & $53.5\pm2.3$ & $80.7\pm23.4$ & $+0.9\pm20.2$ \\*
SFT $-$ Base & $-40.5\pm18.2$ & $-47.6\pm1.4$ & $-32.2\pm21.7$ & $-40.5\pm35.3$ & $-3.9\pm21.8$ \\*
Aligned $-$ Base & $-28.9\pm25.3$ & $-27.9\pm24.4$ & $-46.3\pm2.3$ & $-19.2\pm23.5$ & $+0.9\pm20.1$ \\
\midrule
\multicolumn{6}{@{}l}{\textbf{T\"ulu-3-8B}}\\*
Base (B) & $99.0\pm0.1$ & $98.9\pm0.3$ & $99.1\pm0.2$ & $99.1\pm0.2$ & $-0.0\pm0.2$ \\*
SFT (S) & $81.2\pm6.2$ & $83.9\pm2.0$ & $77.3\pm3.9$ & $77.0\pm5.8$ & $+4.6\pm5.4$ \\*
Aligned (A) & $88.0\pm15.9$ & $80.7\pm19.5$ & $67.5\pm15.4$ & $60.0\pm2.6$ & $+27.1\pm15.1$ \\*
SFT $-$ Base & $-17.9\pm6.3$ & $-15.0\pm2.0$ & $-21.8\pm3.8$ & $-22.1\pm5.8$ & $+4.7\pm5.3$ \\*
Aligned $-$ Base & $-11.0\pm15.9$ & $-18.2\pm19.6$ & $-31.6\pm15.4$ & $-39.1\pm2.4$ & $+27.1\pm15.1$ \\
\midrule
\multicolumn{6}{@{}l}{\textbf{Zephyr-7B}}\\*
Base (B) & $74.5\pm1.6$ & $71.5\pm2.8$ & $72.3\pm1.4$ & $73.6\pm2.3$ & $+1.4\pm3.0$ \\*
SFT (S) & $80.0\pm2.5$ & $76.0\pm1.4$ & $73.2\pm1.6$ & $73.8\pm2.8$ & $+6.8\pm3.2$ \\*
Aligned (A) & $86.4\pm3.0$ & $80.2\pm4.8$ & $78.9\pm3.1$ & $82.7\pm5.3$ & $+5.3\pm1.2$ \\*
SFT $-$ Base & $+5.5\pm2.4$ & $+4.6\pm3.7$ & $+0.9\pm3.0$ & $+0.2\pm3.1$ & $+5.5\pm3.7$ \\*
Aligned $-$ Base & $+11.8\pm4.4$ & $+8.8\pm5.0$ & $+6.6\pm3.3$ & $+9.1\pm3.3$ & $+3.9\pm3.1$ \\
\end{longtable}
\endgroup
\begingroup
\setlength{\LTcapwidth}{\linewidth}
\setlength{\tabcolsep}{4pt}
\small
\begin{longtable}{@{}lrrrrr@{}}
\caption{\textbf{TD: HL40 payoff responses for each lineage.} Entries are mean $\pm$ SD across five runs, averaging secondary parameters within each run. Normalized claims are multiplied by 100; stage differences and fitted declines are in pp. Stage differences match runs by seed ID. The final column fits all four parameter values within each run before taking the mean and SD.}\label{tab:payoff-lineage-td}\\
\toprule
Lineage / stage & $D_1$ & $D_2$ & $D_3$ & $D_4$ & Fitted drop \\
\midrule
\textbf{TD}: $\tau$ & 0.025 & 0.05 & 0.1 & 0.2 & -- \\
\midrule
\endfirsthead
\multicolumn{6}{c}{\tablename\ \thetable{}: TD (continued)}\\
\toprule
Lineage / stage & $D_1$ & $D_2$ & $D_3$ & $D_4$ & Fitted drop \\
\midrule
\textbf{TD}: $\tau$ & 0.025 & 0.05 & 0.1 & 0.2 & -- \\
\midrule
\endhead
\midrule\multicolumn{6}{r}{Continued on next page}\\\endfoot
\bottomrule\endlastfoot
\multicolumn{6}{@{}l}{\textbf{OLMo-3-7B}}\\*
Base (B) & $90.1\pm5.2$ & $90.9\pm2.9$ & $88.5\pm3.1$ & $76.4\pm10.5$ & $+14.6\pm10.0$ \\*
SFT (S) & $98.1\pm1.4$ & $96.9\pm5.3$ & $98.4\pm2.1$ & $98.6\pm1.0$ & $-1.0\pm2.6$ \\*
Aligned (A) & $83.1\pm4.1$ & $83.6\pm9.6$ & $84.1\pm10.2$ & $83.7\pm8.9$ & $-0.4\pm7.4$ \\*
SFT $-$ Base & $+8.1\pm6.2$ & $+6.0\pm6.6$ & $+9.8\pm4.1$ & $+22.2\pm10.2$ & $-15.6\pm11.0$ \\*
Aligned $-$ Base & $-7.0\pm8.3$ & $-7.3\pm11.3$ & $-4.5\pm12.8$ & $+7.2\pm19.0$ & $-15.0\pm15.4$ \\
\midrule
\multicolumn{6}{@{}l}{\textbf{SmolLM2-1.7B}}\\*
Base (B) & $3.6\pm0.4$ & $4.1\pm1.2$ & $4.4\pm0.4$ & $3.9\pm0.4$ & $-0.1\pm0.6$ \\*
SFT (S) & $62.4\pm4.6$ & $62.6\pm5.9$ & $62.6\pm4.7$ & $60.5\pm9.2$ & $+2.0\pm9.2$ \\*
Aligned (A) & $68.3\pm6.1$ & $67.6\pm5.6$ & $65.3\pm4.0$ & $65.9\pm1.5$ & $+2.4\pm2.5$ \\*
SFT $-$ Base & $+58.8\pm4.4$ & $+58.4\pm6.4$ & $+58.2\pm5.0$ & $+56.5\pm9.0$ & $+2.2\pm9.2$ \\*
Aligned $-$ Base & $+64.6\pm6.4$ & $+63.4\pm6.3$ & $+60.9\pm4.3$ & $+61.9\pm1.7$ & $+2.5\pm3.0$ \\
\midrule
\multicolumn{6}{@{}l}{\textbf{DeepSeekMath-7B}}\\*
Base (B) & $30.5\pm2.1$ & $32.1\pm3.8$ & $33.2\pm3.3$ & $32.2\pm2.7$ & $-1.2\pm2.8$ \\*
SFT (S) & $69.0\pm3.9$ & $68.3\pm1.3$ & $69.8\pm3.6$ & $68.2\pm1.2$ & $+0.4\pm3.7$ \\*
Aligned (A) & $68.5\pm2.4$ & $71.1\pm5.1$ & $70.0\pm4.0$ & $67.7\pm5.9$ & $+1.9\pm5.0$ \\*
SFT $-$ Base & $+38.5\pm3.2$ & $+36.2\pm4.1$ & $+36.7\pm4.3$ & $+36.0\pm3.7$ & $+1.7\pm5.1$ \\*
Aligned $-$ Base & $+38.0\pm2.1$ & $+39.0\pm6.6$ & $+36.8\pm4.0$ & $+35.5\pm5.0$ & $+3.1\pm4.6$ \\
\midrule
\multicolumn{6}{@{}l}{\textbf{InternLM2-7B}}\\*
Base (B) & $36.2\pm2.7$ & $35.6\pm3.1$ & $36.8\pm2.2$ & $37.4\pm1.2$ & $-1.5\pm2.1$ \\*
SFT (S) & $63.6\pm0.9$ & $63.0\pm1.4$ & $63.2\pm0.7$ & $62.9\pm1.0$ & $+0.5\pm0.9$ \\*
Aligned (A) & $63.1\pm0.9$ & $62.1\pm0.9$ & $62.2\pm1.8$ & $61.0\pm0.8$ & $+1.8\pm1.2$ \\*
SFT $-$ Base & $+27.5\pm3.3$ & $+27.4\pm2.1$ & $+26.5\pm2.5$ & $+25.6\pm1.4$ & $+2.0\pm2.5$ \\*
Aligned $-$ Base & $+27.0\pm3.4$ & $+26.6\pm3.3$ & $+25.5\pm3.8$ & $+23.7\pm1.3$ & $+3.3\pm3.2$ \\
\midrule
\multicolumn{6}{@{}l}{\textbf{Nemotron-Cascade-8B}}\\*
Base (B) & $58.5\pm7.3$ & $57.0\pm5.2$ & $60.7\pm3.2$ & $56.6\pm13.0$ & $+1.3\pm13.9$ \\*
SFT (S) & $23.0\pm4.6$ & $25.2\pm5.6$ & $25.0\pm4.7$ & $25.5\pm2.5$ & $-1.7\pm2.6$ \\*
Aligned (A) & $25.4\pm3.8$ & $26.4\pm3.3$ & $27.1\pm6.5$ & $23.6\pm2.8$ & $+2.1\pm3.1$ \\*
SFT $-$ Base & $-35.5\pm6.9$ & $-31.8\pm10.3$ & $-35.6\pm3.4$ & $-31.1\pm14.0$ & $-3.0\pm13.0$ \\*
Aligned $-$ Base & $-33.2\pm10.3$ & $-30.6\pm3.6$ & $-33.6\pm7.5$ & $-32.9\pm14.8$ & $+0.8\pm16.5$ \\
\midrule
\multicolumn{6}{@{}l}{\textbf{T\"ulu-3-8B}}\\*
Base (B) & $26.6\pm5.9$ & $27.3\pm3.5$ & $28.7\pm2.6$ & $27.2\pm3.2$ & $-0.5\pm3.4$ \\*
SFT (S) & $83.3\pm3.6$ & $81.4\pm2.8$ & $79.9\pm8.1$ & $86.2\pm4.6$ & $-3.6\pm5.6$ \\*
Aligned (A) & $75.9\pm7.2$ & $71.1\pm5.6$ & $69.6\pm11.7$ & $68.4\pm7.9$ & $+6.0\pm13.3$ \\*
SFT $-$ Base & $+56.7\pm4.9$ & $+54.1\pm5.1$ & $+51.2\pm9.4$ & $+58.9\pm4.2$ & $-3.1\pm3.1$ \\*
Aligned $-$ Base & $+49.3\pm3.5$ & $+43.8\pm5.5$ & $+40.9\pm11.8$ & $+41.2\pm9.2$ & $+6.6\pm11.2$ \\
\midrule
\multicolumn{6}{@{}l}{\textbf{Zephyr-7B}}\\*
Base (B) & $31.6\pm3.1$ & $31.4\pm1.7$ & $32.4\pm1.3$ & $31.9\pm2.8$ & $-0.4\pm2.7$ \\*
SFT (S) & $71.1\pm1.7$ & $71.1\pm1.4$ & $70.1\pm1.8$ & $70.0\pm2.4$ & $+1.2\pm1.8$ \\*
Aligned (A) & $53.9\pm1.4$ & $53.1\pm0.8$ & $51.8\pm2.5$ & $52.3\pm2.2$ & $+1.4\pm2.3$ \\*
SFT $-$ Base & $+39.5\pm4.2$ & $+39.6\pm2.7$ & $+37.7\pm1.9$ & $+38.1\pm3.3$ & $+1.6\pm3.5$ \\*
Aligned $-$ Base & $+22.3\pm4.2$ & $+21.7\pm2.2$ & $+19.4\pm2.6$ & $+20.5\pm2.6$ & $+1.8\pm2.6$ \\
\end{longtable}
\endgroup
\endgroup

\FloatBarrier

\clearpage

\section{A Seven-Stage Nemotron-Cascade Case Study}
\label{app:cascade}
\subsection{Checkpoints, payoff settings, and measurements}
This case study follows one released lineage through seven checkpoints: Qwen3-8B-Base, SFT, RLHF, instruction-following RL (IFRL), mathematical RL (MathRL), code RL (CodeRL), and software-engineering RL (SWE-RL). The canonical Aligned checkpoint for this lineage is \emph{RLHF}. SWE-RL is the endpoint of this longer case study. The primary 21-checkpoint panel's three-stage comparison is complemented here by four subsequent checkpoints.

Base uses \path{Qwen/Qwen3-8B-Base}. SFT through CodeRL use the correspondingly named subfolders of \path{nvidia/Nemotron-Cascade-8B-Intermediate-ckpts}; SWE-RL uses \path{nvidia/Nemotron-Cascade-8B}. Table~\ref{tab:cascade-training-stages} explains the training added at each checkpoint, following the publisher's model cards and training description.\footnote{\href{https://huggingface.co/nvidia/Nemotron-Cascade-8B-Intermediate-ckpts}{NVIDIA intermediate-checkpoint model card}; \href{https://huggingface.co/nvidia/Nemotron-Cascade-8B}{final model card}; \href{https://arxiv.org/html/2512.13607v1}{training description, Sections 3--4}.} These checkpoints trace successive training stages within the same model lineage. All checkpoints use the same game prompt template and action-label interface, bfloat16 inference, temperature 0.7, top-$p$ 0.8, and top-$k$ 20.

\begin{table}[!htbp]
\centering\small
\caption{\textbf{What changes across the seven Nemotron-Cascade checkpoints?} Rows follow the released training sequence. Base, SFT, and RLHF are the three stages used for this lineage in the main panel; the remaining rows extend that sequence.}
\label{tab:cascade-training-stages}
\begin{tabularx}{\linewidth}{@{}p{0.14\linewidth}X@{}}
\toprule
Checkpoint & Training completed at this step \\
\midrule
Base & Qwen3-8B pretrained model, before Nemotron-Cascade post-training. \\
SFT & Multi-stage supervised fine-tuning on example responses spanning general dialogue, reasoning, coding, and other domains. \\
RLHF & Reinforcement learning using a learned human-preference reward. This is our main panel's Aligned checkpoint. \\
IFRL & Instruction-following reinforcement learning, rewarding adherence to explicit response constraints; applied in non-thinking mode for this unified model. \\
MathRL & Reinforcement learning on mathematical problems, using answer-correctness rewards. \\
CodeRL & Reinforcement learning on programming problems, using code-execution tests as rewards. \\
SWE-RL & Software-engineering reinforcement learning for code repair, using an execution-free reward that compares proposed repairs with reference patches. \\
\bottomrule
\end{tabularx}
\end{table}

All seven checkpoints are evaluated on the same original 82 configurations. Relative to the main panel in Appendix~\ref{app:games}, PGG includes MPCR $0.7$, giving $\{0.4,0.5,0.6,0.7,0.8\}$, and CPR uses $\theta\in\{0.2,0.5,0.8\}$ instead of $\{0.2,0.4,0.6,0.8\}$; all other parameters are unchanged. Thus PGG has ten complete configurations and CPR twelve, compared with eight and sixteen in the main panel. Each checkpoint has 20 one-shot draw IDs per configuration and five trajectory seeds at each of HL2 and HL40; repeated rates average all 200 logged rounds and all seats within configuration. All comparisons in this appendix use this common 82-configuration grid.

We compare cooperation rates and payoff responses on the same five games: PGG, CPR, Donation-PD, Volunteer, and Assurance. A separate history table adds Trust to match the six-binary-game average in the main text. For each checkpoint and run, we first average cooperation over configurations within a game, then average games equally. For payoff responses, we fit the main parameter while holding secondary parameters fixed, then average these fits and games (Equation~\ref{eq:S}). PGG therefore fits five points here and CPR three. Assurance uses the $\lambda$ span $0.35$; fitting successful payoff $h=3(1+\lambda)$ over span $1.05$ gives the identical fitted drop.

The result tables report mean $\pm$ sample SD across 20 one-shot draws or five repeated-game runs \emph{within a checkpoint}. For history and consecutive-stage comparisons, we subtract results with the same seed ID before computing the SD. This SD describes variation across runs of this one lineage. OS denotes one-shot decisions. As in Equation~\ref{eq:history}, $H_s$ is HL40 minus HL2, and $K_{s/B}$ is that history change minus Base's; both are reported in pp. Negative $K_{s/B}$ means a larger cooperation decline than Base's. Rows follow the released training sequence.

\subsection{Higher cooperation with a smaller fitted payoff decline}
\begin{figure}[!htbp]
\centering
\includegraphics[width=\linewidth]{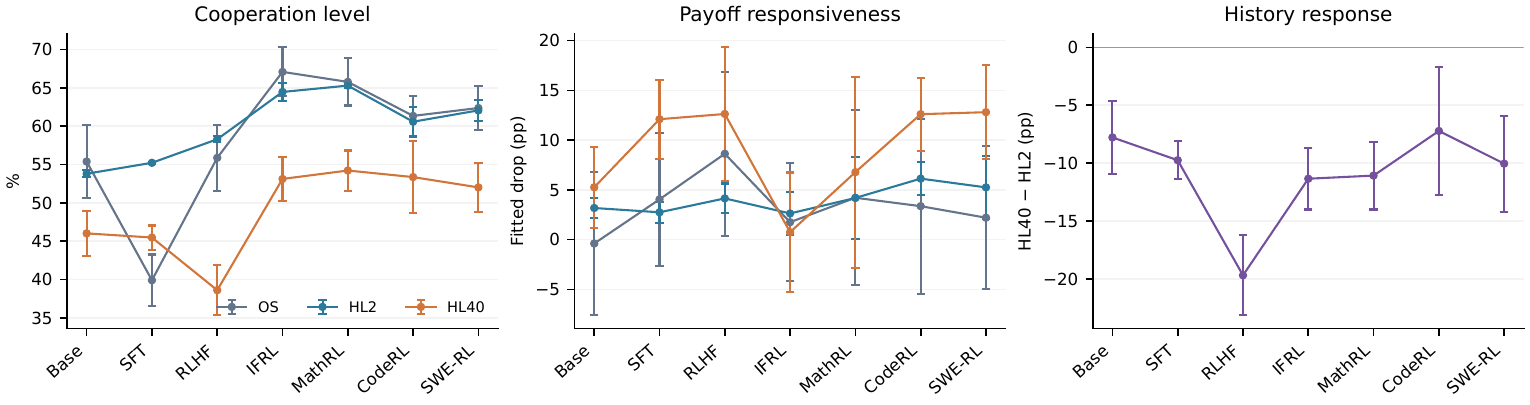}
\caption{The same five symmetric binary games across seven checkpoints: cooperation level, all-point fitted payoff responsiveness, and HL40--HL2 history response. Points and whiskers show mean $\pm$ within-checkpoint replicate SD (20 one-shot draws or five trajectory seeds); the history contrast pairs the five seeds. Each panel has its own labeled metric and axis. RLHF is the canonical Aligned endpoint.}
\label{fig:cascade-common}
\end{figure}

Figure~\ref{fig:cascade-common} and Tables~\ref{tab:cascade-common_binary5-levels}--\ref{tab:cascade-common_binary5-responses} compare the cooperation rate and fitted payoff drop across these same five games. From RLHF to IFRL, mean cooperation rises by 11.22 pp in one-shot decisions, 6.18 pp at HL2, and 14.53 pp at HL40. At HL40, the fitted payoff drop falls from 12.63 to 0.75 pp: cooperation becomes higher but changes less across the tested payoff range. The same combination occurs in one-shot and HL2 evaluations. This transition combines higher cooperation with a smaller fitted response to payoff changes. The figure displays the training path; the tables give means, SDs, and paired history changes.
\begin{table}[!htbp]
\centering\small
\setlength{\tabcolsep}{4pt}
\caption{Same five symmetric binary games for both level and response: PGG, CPR, Donation-PD, Volunteer, Assurance. Levels are percentages; history contrasts are pp. Mean $\pm$ SD across replicates within each checkpoint. Repeated contrasts pair five nominal seeds.}\label{tab:cascade-common_binary5-levels}
\begin{tabular}{lrrrrr}\toprule
Stage & OS & HL2 & HL40 & $H$ & $K_{s/B}$\\\midrule
Base & $55.39\pm4.76$ & $53.80\pm0.48$ & $46.01\pm2.95$ & $-7.79\pm3.15$ & $+0.00\pm0.00$ \\
SFT & $39.89\pm3.36$ & $55.23\pm0.16$ & $45.47\pm1.57$ & $-9.76\pm1.66$ & $-1.97\pm4.53$ \\
RLHF & $55.88\pm4.31$ & $58.30\pm0.41$ & $38.59\pm3.27$ & $-19.71\pm3.45$ & $-11.92\pm5.32$ \\
IFRL & $67.10\pm3.21$ & $64.48\pm1.18$ & $53.12\pm2.87$ & $-11.36\pm2.66$ & $-3.57\pm4.61$ \\
MathRL & $65.79\pm3.09$ & $65.31\pm0.35$ & $54.22\pm2.62$ & $-11.09\pm2.93$ & $-3.30\pm1.97$ \\
CodeRL & $61.35\pm2.62$ & $60.59\pm1.93$ & $53.36\pm4.74$ & $-7.23\pm5.56$ & $+0.56\pm6.12$ \\
SWE-RL & $62.38\pm2.93$ & $62.07\pm1.39$ & $52.01\pm3.21$ & $-10.05\pm4.15$ & $-2.27\pm2.96$ \\
\bottomrule
\end{tabular}\end{table}

\begin{table}[!htbp]
\centering\small
\setlength{\tabcolsep}{4pt}
\caption{Same five symmetric binary games for both level and response: PGG, CPR, Donation-PD, Volunteer, Assurance. All-point fitted responses in pp; mean $\pm$ replicate SD (20 draws for OS, five seeds for HL2/HL40).}\label{tab:cascade-common_binary5-responses}
\begin{tabular}{lrrr}\toprule
Stage & OS & HL2 & HL40\\\midrule
Base & $-0.39\pm7.21$ & $+3.18\pm1.01$ & $+5.26\pm4.07$ \\
SFT & $+4.05\pm6.67$ & $+2.75\pm1.04$ & $+12.11\pm3.96$ \\
RLHF & $+8.63\pm8.25$ & $+4.15\pm1.50$ & $+12.63\pm6.70$ \\
IFRL & $+1.76\pm5.90$ & $+2.64\pm2.19$ & $+0.75\pm6.00$ \\
MathRL & $+4.21\pm8.78$ & $+4.19\pm4.10$ & $+6.79\pm9.61$ \\
CodeRL & $+3.37\pm8.79$ & $+6.13\pm1.68$ & $+12.61\pm3.66$ \\
SWE-RL & $+2.21\pm7.16$ & $+5.25\pm3.14$ & $+12.81\pm4.74$ \\
\bottomrule
\end{tabular}\end{table}

\FloatBarrier

\subsection{History responses across game sets}
On the six-binary-game panel, every post-training checkpoint has a more negative mean history response than Base (Table~\ref{tab:cascade-legacy-levels}). The Base and SWE-RL HL40 levels are $54.46\pm2.46\%$ and $54.27\pm2.36\%$, while their history changes are $-6.43\pm2.60$ and $-10.43\pm3.04$ pp. Similar endpoint levels thus coexist with different responses to history. Intermediate HL40 levels range from RLHF's $38.73\%$ to MathRL's $57.50\%$.

The five-game panel supplies the like-for-like level/response comparison above. In that panel, five of six post-training history contrasts are more negative than Base; CodeRL has $K_{s/B}=+0.56\pm6.12$ pp. The six-binary-game panel, which includes Trust, has negative Base-relative history contrasts at all six post-training checkpoints. Both summaries specify their game sets.
\begin{table}[!htbp]
\centering\small
\setlength{\tabcolsep}{4pt}
\caption{Six-binary-game cooperation and history panel, including Trust and excluding TD. Levels are percentages; history contrasts are pp. Mean $\pm$ SD across replicates within each checkpoint. Repeated contrasts pair five nominal seeds.}\label{tab:cascade-legacy-levels}
\begin{tabular}{lrrrrr}\toprule
Stage & OS & HL2 & HL40 & $H$ & $K_{s/B}$\\\midrule
Base & $57.82\pm4.91$ & $60.89\pm0.45$ & $54.46\pm2.46$ & $-6.43\pm2.60$ & $+0.00\pm0.00$ \\
SFT & $38.96\pm3.60$ & $56.44\pm0.28$ & $45.84\pm1.51$ & $-10.60\pm1.56$ & $-4.17\pm3.84$ \\
RLHF & $50.59\pm3.83$ & $58.58\pm0.50$ & $38.73\pm2.81$ & $-19.85\pm3.02$ & $-13.42\pm4.61$ \\
IFRL & $66.08\pm3.42$ & $67.63\pm0.72$ & $55.92\pm2.06$ & $-11.70\pm1.90$ & $-5.27\pm3.50$ \\
MathRL & $63.39\pm3.04$ & $69.10\pm0.42$ & $57.50\pm2.19$ & $-11.60\pm2.26$ & $-5.17\pm2.05$ \\
CodeRL & $59.92\pm2.79$ & $63.06\pm1.05$ & $55.53\pm4.01$ & $-7.53\pm4.58$ & $-1.09\pm5.12$ \\
SWE-RL & $60.36\pm2.88$ & $64.70\pm1.02$ & $54.27\pm2.36$ & $-10.43\pm3.04$ & $-4.00\pm1.95$ \\
\bottomrule
\end{tabular}\end{table}

\FloatBarrier

\normalsize
\FloatBarrier

\end{document}